\documentclass[aps,pre,twocolumn,superscriptaddress,showpacs,longbibliography]{revtex4-1}
\usepackage{epsfig}
\usepackage{amsmath,amssymb,bm}
\usepackage{graphics}
\usepackage{epsfig}
\usepackage{graphicx}
\usepackage{verbatim}
\usepackage{makecell}

\usepackage{color}

\usepackage{multirow}
\newcommand{\spc}[1]{ \hspace{#1 em}}

\begin{document}
\title{Complex distributions emerging in filtering and compression  
}

\author{G.~J.~Baxter}

\affiliation{Departamento de F{\'\i}sica da Universidade de Aveiro $\&$ I3N, Campus Universit\'ario de Santiago, 3810-193 Aveiro, Portugal}

\author{R.~A.~da~Costa}

\affiliation{Departamento de F{\'\i}sica da Universidade de Aveiro $\&$ I3N, Campus Universit\'ario de Santiago, 3810-193 Aveiro, Portugal}

\author{S.~N. Dorogovtsev}

\affiliation{Departamento de F{\'\i}sica da Universidade de Aveiro $\&$ I3N, Campus Universit\'ario de Santiago, 3810-193 Aveiro, Portugal}
\affiliation{A.F. Ioffe Physico-Technical Institute, 194021 St. Petersburg, Russia}

\author{J.~F.~F. Mendes}
\affiliation{Departamento de F{\'\i}sica da Universidade de Aveiro $\&$ I3N, Campus Universit\'ario de Santiago, 3810-193 Aveiro, Portugal}
\affiliation{School of Computer and Communication Sciences and School of Life Sciences, \'Ecole Polytechnique F\'ed\'eral de Lausanne, 1015 Lausanne EPFL, Switzerland}

\begin{abstract}
\noindent
In filtering, each output is produced by a certain number of different inputs. We explore the 
statistics of this degeneracy in an explicitly treatable filtering problem in which filtering performs the maximal compression of relevant information contained in inputs (arrays of zeroes and ones). 
This problem serves as a reference model for the statistics of filtering and related sampling problems. 
The filter patterns in this problem conveniently allow a microscopic, combinatorial consideration. This allows us to find the statistics of outputs, namely the exact distribution of output degeneracies, for 
arbitrary input sizes. We observe that the resulting degeneracy distribution of outputs decays as $e^{-c\log^\alpha \!d}$ with degeneracy $d$, where $c$ is a constant and exponent $\alpha>1$, i.e. faster than a power law. 
Importantly, its form 
essentially depends on the size of the input data set, appearing to be closer to a power-law dependence for small data set sizes than for large ones. We demonstrate that for sufficiently small input data set sizes 
typical for empirical studies, this distribution could be easily perceived as a power law. 
We extend our results to filter patterns of various sizes and demonstrate that the shortest filter pattern provides the maximum informative representations of the inputs.
\end{abstract}
\maketitle


\section{Introduction}
\label{s1} 

Compression, filtering, and cryptography are related areas in signal and information processing 
\cite{borda2011fundamentals,hankerson2003introduction,tishby2000information,gehani2003dna}.
By definition, a large number of possible inputs are mapped to a smaller number of possible outputs, so that a given output may correspond to multiple inputs, which number is the output's degeneracy.
A similar problem emerges in cooperative systems with a large number of local minima in the energy landscape, in particular, in spin glasses and deep learning neural networks \cite{hartmann2005phase,mezard1987spin,keskar2016large}. 
The configuration space of a system of this sort can be divided into a set of domains (basins) of attraction of these minima. 
One can ask: what is the statistics of these domains of attraction, what is the distribution of their sizes? This is analogous to the degeneracy statistics problem.
These statistics are important because the degeneracy distribution gives the most relevant entropy when studying information problems \cite{baek2011zipf}, that is, it is the distribution containing the most information relevant to the system under study. 

These issues were explored in a recent series of works \cite{song2017emergence,cubero2018minimum,haimovici2015criticality,cubero2018minimally,song2018resolution} which exploited the principle of maximum entropy,  
see also Refs.~\cite{bianconi2009entropy,bianconi2007statistical,bashkirov2000information,dover2004short,visser2013zipf}. 
The finding of Refs.~\cite{song2017emergence,cubero2018minimum,haimovici2015criticality,cubero2018minimally,song2018resolution} is that 
maximally informative samples drawn from data exhibit statistics with broad distributions. 
More specifically, their entropy optimization based theory predicts power-law like distributions of degeneracy of maximally informative outputs (minimal sufficient representations). The title of the paper ``Minimum description codes are critical'' \cite{cubero2018minimum} published in the journal Entropy highlights the message of that theory in the most succinct way. 
On the other hand, distributions markedly distinct from power laws can be observed in empirical distributions for this kind of problem, for example, a collection of empirical curves in Fig.~1 of Ref.~\cite{cubero2018minimum}. 

\begin{figure}[t]
\begin{center}
\includegraphics[scale=0.15]{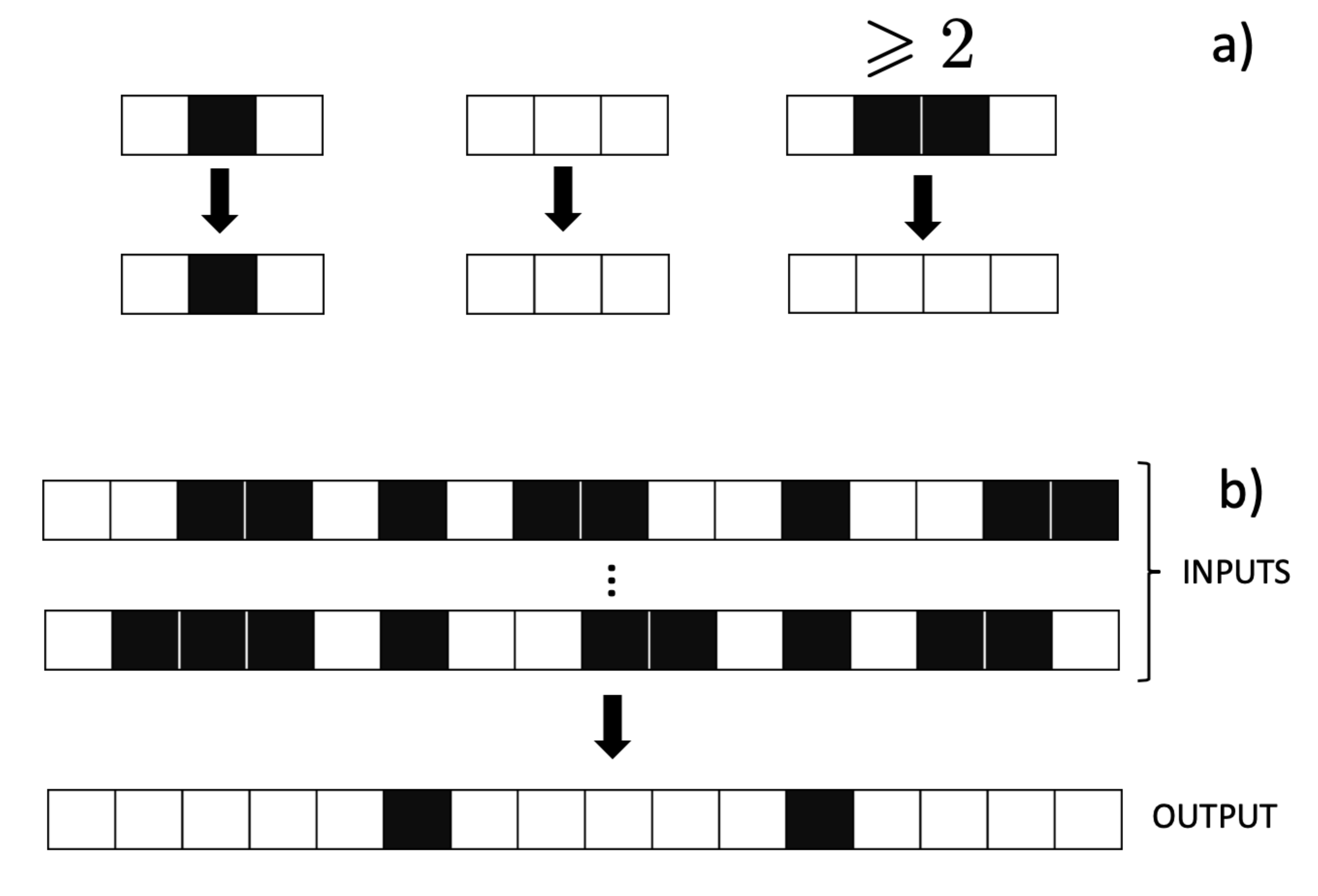}
\end{center}
\caption{
(a) Filter extracting single ones and their positions from sequences of ones and zeroes. Here the filter pattern is a single one. Ones and zeroes are shown as black and white pixels, respectively. (b) A number of different inputs produce the same output. This number is the degeneracy of the output. 
}
\label{f0}       
\end{figure}

One obstacle to understanding such phenomena is that the need to obtain sufficient samples to resolve the degeneracy distribution limits the system size which can be examined. In the present work we contribute to these efforts by proposing a model for which sampling distributions can be obtained exactly, for the entire state space and up to large system sizes.
 We introduce and explore a reference filtering problem straightforwardly treatable through purely combinatorial techniques.
 This filter 
 extracts all positions of a given local pattern in the input, see Fig.~\ref{f0}.
We study a family of such filter patterns and demonstrate that the smallest pattern, extracting single ones from the inputs, generates outputs with the highest entropy of the degeneracy distribution, called relevance, Fig. \ref{fig3} and Table \ref{su40}. 
According to Refs.~\cite{cubero2018minimum,cubero2018minimally}, this type of information entropy is maximized by the representations of the {samples of a complex system} which provide the most information, the maximally informative representations, { which occurs in a critical region where the sample distribution becomes complex. Outside of this region outputs are either nearly all the same, or nearly all different, giving no useful information about the underlying system. Note that the entropy of the samples themselves, called the resolution, is not maximum in this critical region, but in the limit that all samples are different.}
{We establish} 
a link between filtering and sampling problems based on the idea that filter patterns of different lengths with their positions can serve as sampling variables. {We demonstrate that the behavior of our filtering problem is analogous to that found in the more complex systems of Refs. \cite{song2017emergence,cubero2018minimum,haimovici2015criticality,cubero2018minimally,song2018resolution}. The entropy of the degeneracy distribution of filter outputs increases as the filter length decreases, but vanishes in the extreme limit, for which resolution is maximised, with outputs being identical to inputs. Thus within a family of filters, the most informative filter is the shortest one with length greater than one.}

We directly obtain the complex degeneracy distributions for outputs generated from various input sets. We develop efficient recursive methods that allow us to find this distribution for large input strings, as can be seen in Fig.\ref{f1}, which would not be accessible through empirical sampling methods.
Due to the tractability of this problem, we are able to identify precisely how these distributions deviate from a power-law dependence. We find scaling forms that accurately describe the tail of the degeneracy distributions (see Fig.~\ref{f4p}).  
We discover that these distributions are essentially shaped by the size of the input data. That is, input data sets of relatively small size, typical for empirical studies, produce degeneracy distributions of outputs that are closer to a power law than the distributions of outputs from very large input data sets. 
Finally, we develop a mean field theory and obtain the asymptotics of the degeneracy distribution and the spectrum of degeneracies.  
Our findings indicate that the phenomena we observe should apply to more general filtering and compression problems.

Our paper is organized in the following way. In Sec.~\ref{s2} we introduce our filter and the synthetic input data sets for it. 
{In Sec.~\ref{a4} we describe how the filtering problem can be viewed as sampling of a complex system.}
In Secs.~\ref{s3}, \ref{s3p}, and \ref{s3pp} we obtain the basic relations for the degeneracy of outputs, develop our algorithm, obtain the complex degeneracy distributions of outputs for the complete input data sets, and describe their features. We develop the mean-field theory of these distributions in filtering in Sec.~\ref{s3ppp}. In Sec.~\ref{s4} we obtain the degeneracy distributions of outputs for uniformly random input data sets of various sizes. In Sec.~\ref{s5} we discuss our results and indicate possible generalizations.
In the Appendix we 
give the combinatorial derivations of recursive relations used in Sec.~\ref{s3} and explicit asymptotics. 
The exact results obtained by our algorithm and recursions are provided in the Supplementary Material \cite{supplementary}. 


\section{A reference filtering problem}
\label{s2} 

We study the distribution of outputs in a solvable filtering problem by implementing a purely combinatorial, microscopic approach not involving entropy considerations. 
Let the input data be a set of $N$ strings of zeroes
 and ones 
$(x_i)$, $x_i=0,1$,  of length $n$, 
 assuming the periodic condition $x_1=x_{n+1
}$. 
We consider two types of data set. The first set is the complete set of all possible unique inputs. Its size $N$ is determined by the size $n$ of inputs, $N = 2^n$.
Second, we consider data sets of arbitrary size $N$ consisting of strings of uniformly randomly generated zeroes 
 and ones 
 constrained by the same periodic condition as above.
 In the latter situation, some of the elements of a data set may coincide. Clearly, in the limit $N\to\infty$, we arrive at a situation equivalent to the complete data set. (We stress that the random data set of size $N=2^n$ differs from the the complete data set.) 

\begin{figure*}[t]
\begin{center}
\includegraphics[scale=0.5]{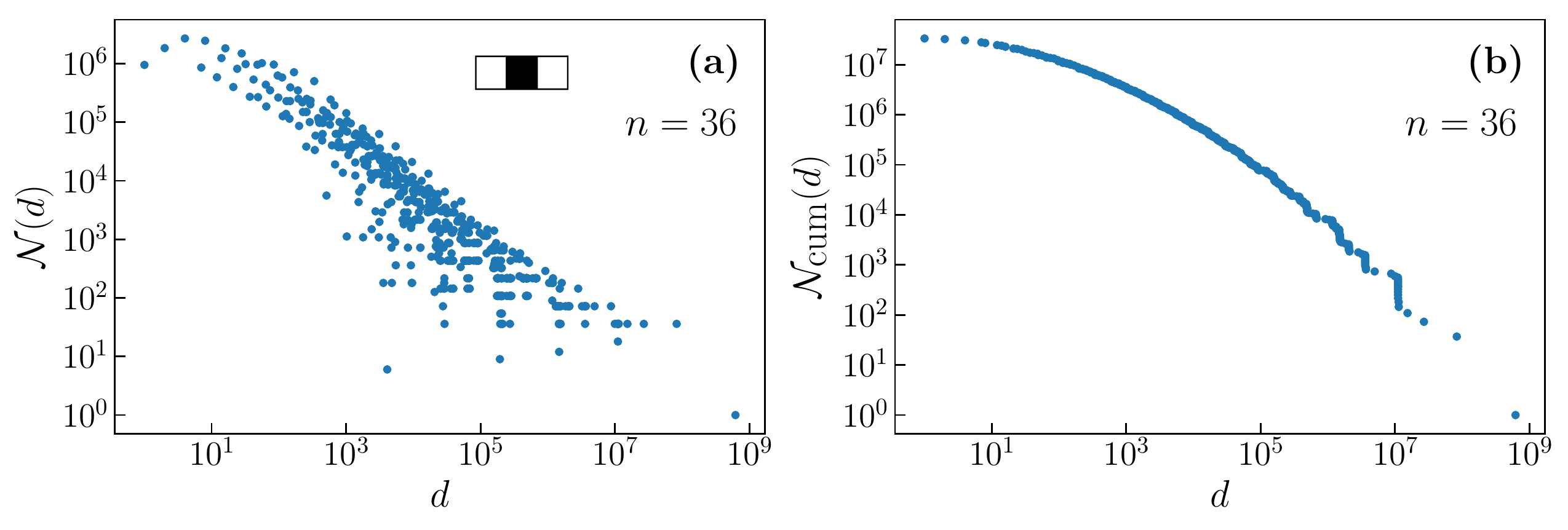}
\includegraphics[scale=0.5]{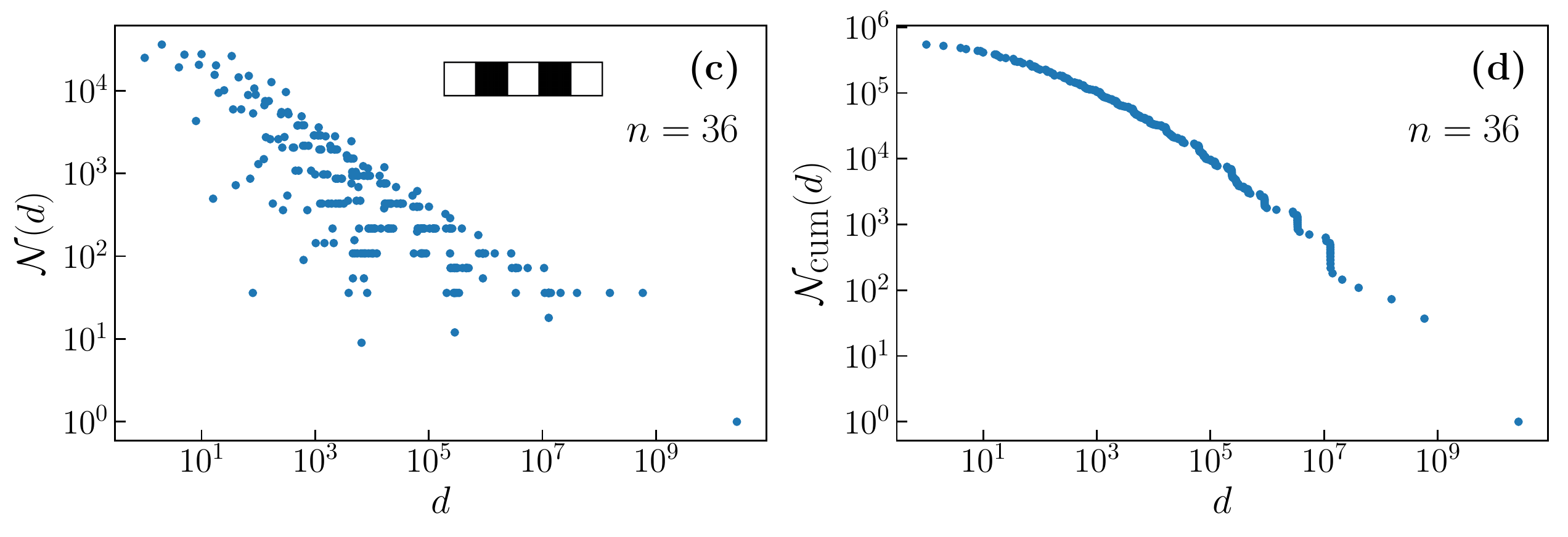}
\includegraphics[scale=0.5]{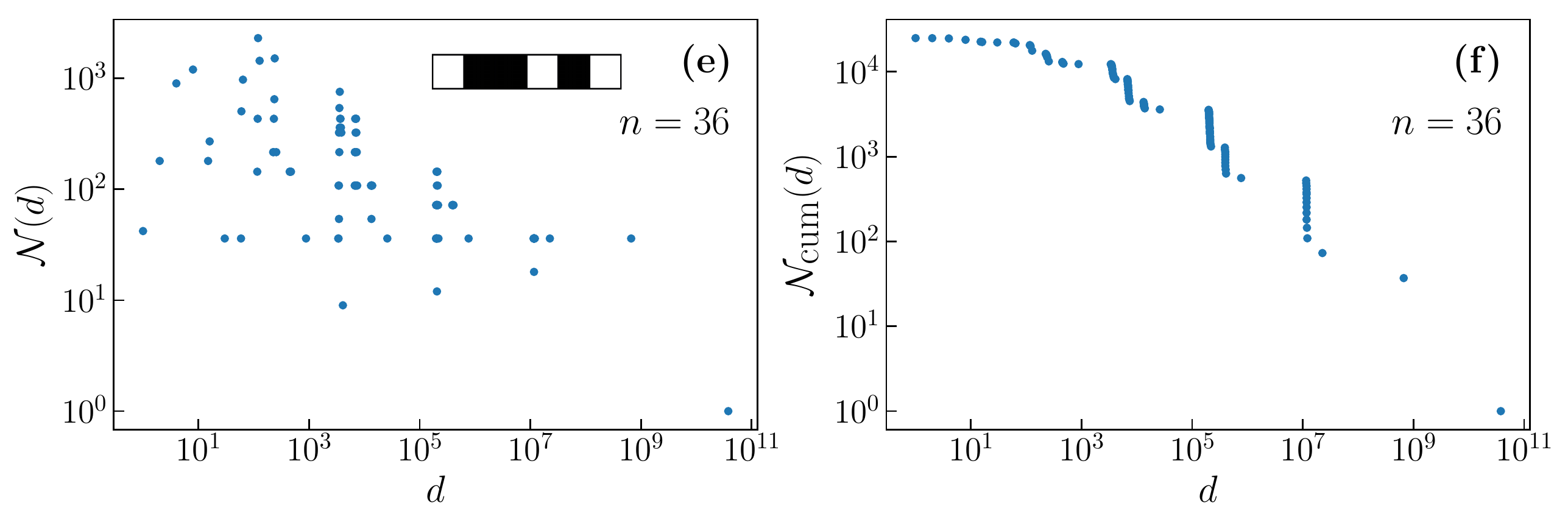}
\includegraphics[scale=0.5]{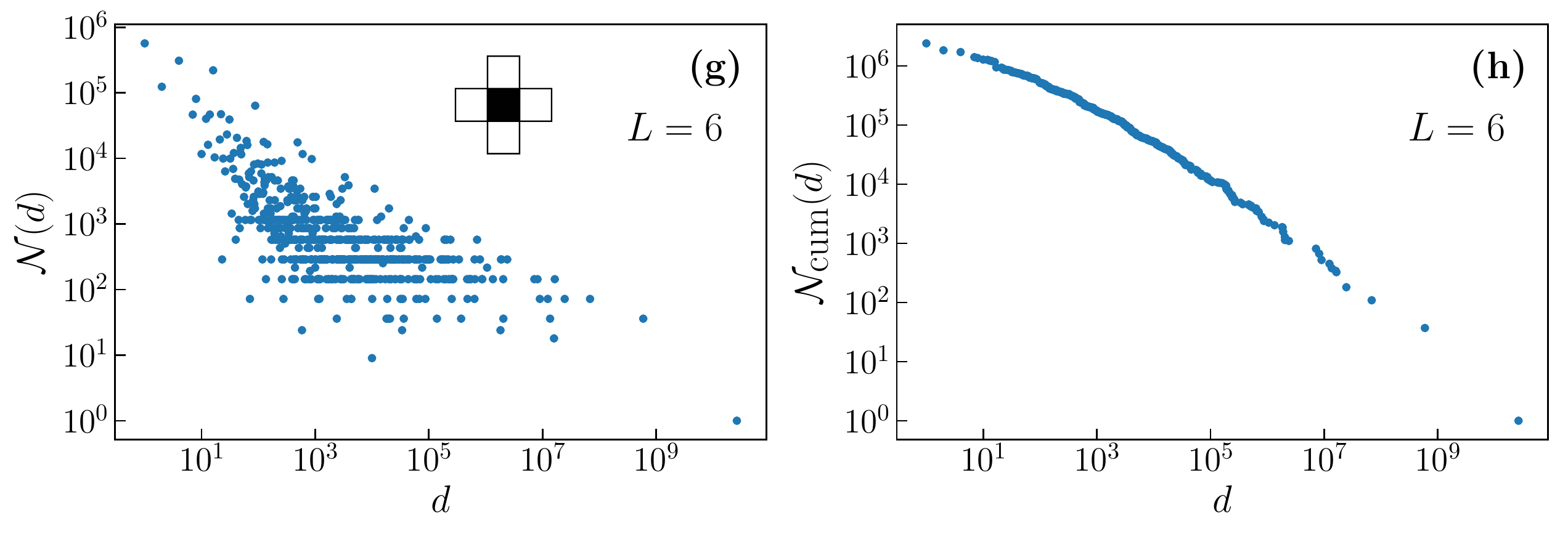}
\end{center}
\caption{
Degeneracy distribution and cumulative degeneracy distributions for alternative filters to the one used in our main results.
(a, b) Filter pattern $01010$ for binary input strings of length $n=36$. 
(c, d) Filter pattern $01010$ for binary input strings of length $n=36$. 
(e, f) Filter pattern $011010$ for binary input strings of length $n=36$. 
(g, h) Filter pattern (as illustrated in figure) for 2D binary input grid of side length $L=6$ ($n=36$ entries total).
}
\label{fig3}       
\end{figure*}

The filter works as follows: every instance of a specific pattern in the input is marked by a one in the corresponding position in the output. All other positions are marked with zeroes. This produces a minimal coding of the positions of the pattern occurrences in the input.
We illustrate the results from a few example filter patterns in Fig. \ref{fig3}. We observe complex degeneracy distributions reminiscent of those observed in, for example, Ref. \cite{marsili2013sampling}.

For the sake of simplicity, in the remainder of this paper we focus on the simplest of these filters.
Each sequence of ones of length $1$ in the input (i.e., every $1$ whose neighbors are both zeroes)
 gives $1$ at the same position in the output. All other sequences of ones 
 or zeroes in the input produce zeroes in the corresponding places in the output, as shown in Fig. \ref{f0}. In other words, the input vector $(x_i)$, $i=1,2,...,n$, $x_i=0,1$, is transformed to the output vector with the following components: 
\begin{equation}
y_i = (1-x_{i-1})x_i(1-x_{i+1}) 
.
\label{10}
\end{equation}
{As we will see, this filter is the maximally informative one, in a family of similar filters.}
Other filter patterns may be analysed using the same methods we describe below.

\section{Filtering as a sampling problem}
\label{a4}
{
Let us discuss the link between filtering and sampling in more detail. 
The input $x_i$ to our filtering problem is a string of $n$ ones and zeroes. It contains alternating contiguous chains of ones and zeroes of different lengths, so that one may completely define the sequence by listing the starting positions of all the strings of ones and their respective lengths.
With this in mind, one may reformulate the problem of filtering for patterns in sequences as sampling 
a system of correlated discrete variables in the following way.
Let us introduce the set of discrete variables $\{s_i^l\}$, for $i=1,...,n$ and $l=1,...,n-1$. The (`spin') variable $s_i^l$ is $1$ if the input has a string of $l$ ones beginning at position $i$, and is $0$ otherwise.
These variables correspond to the outputs of filters for chains of consecutive ones.
We include one more variable which distinguishes between the states of all zeroes and all ones.
The system, then, has $2^m$, $m=n(n-1)+1$, possible states.

Strings of ones in the input do not overlap with each other, which restricts the configurations of the variables $\{s_i^l\}$ that can occur. We include these constraints in our system by imposing a cost for each violation of the mutual exclusion rules, the cost function being 
\begin{equation}
\label{cost_fn}
U(\vec{s}) = 
\sum_{s_i^l,s_j^m} s_i^l s_j^m \left[
  \Theta(i+l,j) + \Theta(j+m,i)
  \right]
,
\end{equation}
where 
\begin{align}
  \Theta(a,b)
  &= \begin{cases} 
1 & a \mod n \geq b, \\
0 & \mbox{otherwise}
\end{cases}
\end{align}
functions as a cyclic Heaviside function.
We now have a complex system consisting of $m$ state variables, whose interactions are governed by the cost function $U$.
Each of the $2^n$ possible input sequences corresponds to a different minima of this cost function.

Now let us consider the sampling of this system. Suppose we may 
measure only $n$ of these variables $s_i^l$. Which $n$ variables will be most informative about the state of the system?
intuitively, shorter filter patterns, {\em i.e.} smaller values of $l$, should provide more detailed information about the system, and indeed the variables $s_i^{l=1}$ contain the most information about the state of the system. 
(To see this, note that the variable $s_i^l$ gives information about the state of $l+2$ input digits. In particular, when $s_i^l=1$, it fully determines $l+2$ digits of the input, which happens for a fraction $2^{-(l+2)}$ of the inputs.
Thus, the total amount of information conveyed by $s_i^l$ in terms of the number of digits of the input it reveals is $(l+2)2^{-(l+2)}$, which is indeed maximum for $l=1$.)

The set $\{s_i^1\}$ is also the most informative representation of the input from the perspective of Ref.~\cite{marsili2013sampling}, in the sense that samples obtained from observing these $n$ variables have a larger entropy of their degeneracy distributions, Eq. (\ref{entropy}), than samples using any other subset of $n$ variables. 
This is apparent from the fact that any subset of $n$ variables containing larger filters will have less available configurations, see also Table \ref{su40}. We detail this analogous behavior further in Sec. \ref{s3pp}.
}


\section{Outputs and their degeneracies for complete input dataset}
\label{s3} 

\begin{figure*}[t]
\begin{center}
\includegraphics[scale=0.5]{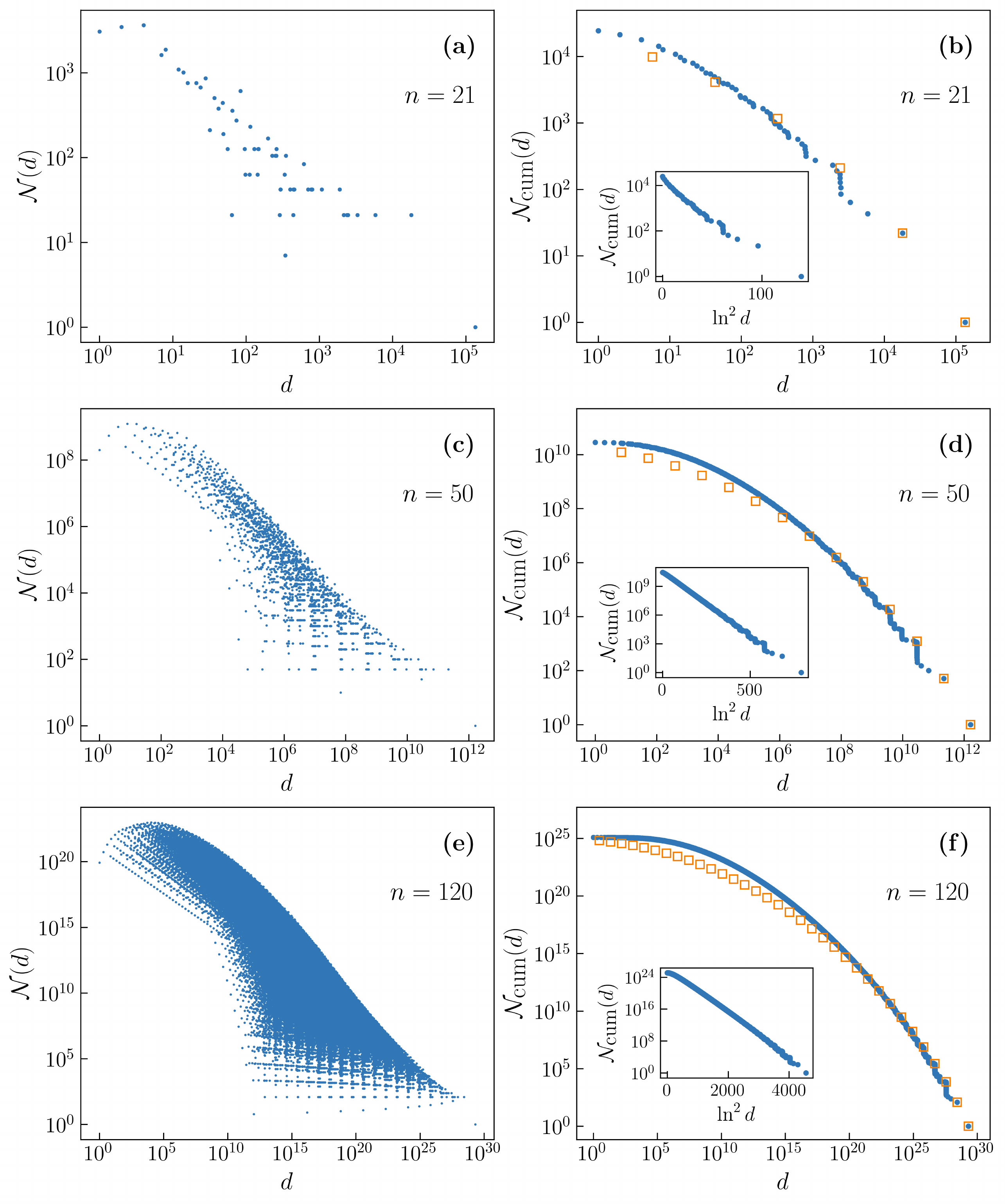}
\end{center}
\caption{
(a,c,e) Degeneracy distribution for the complete input data set: number of outputs of a given degeneracy vs. degeneracy. 
(b,d,f) Cumulative degeneracy distribution for the complete input data set: number of outputs of degeneracy greater than or equal to a given degeneracy vs. degeneracy. 
The input length is $n=21,50,120$. Open square symbols represent a mean-field approximation to the cumulative distribution, given by Eq. (\ref{5000}).
}
\label{f1}       
\end{figure*}

Let us begin by considering the complete input data set, all $2^{n}$ 
different configurations of zeroes and ones. 
Each output consists of isolated ones separated by strings of zeroes of various lengths. The total number of possible outputs, $M(n)$, is then the number of ways of arranging isolated ones in a chain of length $n$.
This number coincides with the number of different configurations of dimers in a closed chain (ring) of length $n$: 
\begin{equation}
M(n) = n \sum_{k=0}^{[n/2]}\frac{1}{n - k} {n - k \choose k}
,
\label{20}
\end{equation}
where $[n/2]$ is the integer part of $n/2$, see Appendix~\ref{a11}. 

Comparing this number for successive values of $n$, we see that 
$M (n) = M (n - 1) + M(n - 2)$, with initial conditions $M (3) = 4$, $M (4) = 7$.
The elements of the sequence may be written in terms of the roots of the characteristic equation \cite{hoggatt1969fibonacci,graham1994concrete,koshy2019fibonacci}
\begin{equation}
z^2 - z = 1 
.
\label{40}
\end{equation}
Thus
\begin{equation}
M(n) = \Bigl(\frac{1 + \sqrt{5}}{2}\Bigr)^n + \Bigl(\frac{1 - \sqrt{5}}{2}\Bigr)^n \cong \Bigl(\frac{1 + \sqrt{5}}{2}\Bigr)^n =z_g^n 
, 
\label{30}
\end{equation}
where the last expression gives the large $n$ limit. 
Here the largest root $(1 + \sqrt{5})/2 = 1.61803... \equiv z_g$ 
is the famous golden ratio. 

The key point for our study is that the degeneracy of an output is the product of the degeneracies of the strings of zeroes 
 between ones in the output. 
 As is clear from Eq.~(\ref{10}), the presence of a $1$ in an output at position $i$ fixes the digits of its inputs at positions $i-1$, $i$, and $i+1$. These digits must be $0$, $1$, and $0$, respectively. This is true for each $1$ in the output.
On the other hand, each of the remaining digits of these inputs compatible with a given output must be either a $0$ or be a $1$ with one or two neighboring ones.  
All the degrees of freedom in the input corresponding to a given output 
correspond to these digits. 

Let an output with $m\geq 1$ ones contain $m$ strings of zeroes with lengths $\ell_1,\ell_2,...,\ell_m$. Then the degeneracy of this output equals 
\begin{equation}
d = \prod_{i=1}^{m}  \tilde{d}(\ell_i) 
. 
\label{41}
\end{equation}
Here $\tilde{d}(\ell)$ is the number of input strings of length $\ell$, having the first and last digits $0$, that  generate an output string of $\ell$ zeroes. 
This number plays an important role in our problem, similar to prime numbers in number theory, so we call the $\tilde{d}(\ell)$ {\em prime degeneracies}. 
Suppose that the output contains $\mu_\ell$ strings of zeroes of length $\ell$, $\ell=1,2,...$, where  
\begin{equation}
m + \sum_{\ell\geq1} \ell \mu_\ell = n
.
\label{42}
\end{equation}
Then Eq.~(\ref{41}) may be rewritten 
\begin{equation}
d = \prod_{\ell\geq 1} [\tilde{d}(\ell)]^{\,\mu_\ell}
\label{43}
\end{equation}
for $m\geq1$.

Let us find the prime degeneracies explicitly. 
The number of these configurations, i.e., the degeneracy $\tilde{d}(\ell)$ of the output string of $\ell$ zeroes, 
can be obtained recursively by taking into account three points: 

(i) Relevant input configurations of length $\ell$ are obtained by inserting 
$0$ or 
$1$ into each relevant configuration of length $\ell-1$ between the first and second positions of the sequence. (Recall that the first and last positions of the input sequence are fixed to 0.) 

(ii) Input strings of length $\ell$ beginning and/or ending with $010$ are irrelevant, and so they should be removed from the set generated at the previous step.  
These configurations can be obtained by inserting two digits $10$ into each relevant input string of length $\ell-2$ between its first and second positions.

(iii) Finally, there exist input strings, compatible with the output string of $\ell$ zeroes, 
that cannot be obtained by inserting a single digit into relevant input strings of length $\ell-1$ between their first and second positions. These are the input strings of length $\ell$ beginning with $0110$. These inputs can be obtained by inserting $110$ into each relevant input string of length $\ell-3$ between their first and second positions.

Following these rules, the degeneracy of a string of $\ell$ zeroes at the output, prime degeneracy $\tilde{d}(\ell)$, can be written recursively as a linear difference equation: 
\begin{equation}
  \tilde{d}(\ell) = 2 \tilde{d}(\ell-1) - \tilde{d}(\ell-2) + \tilde{d}(\ell-3)
\label{eq10}
\end{equation} 
with the initial condition $\tilde{d}(1)=\tilde{d}(2)=\tilde{d}(3)=1$. 
The solution of Eq.~(\ref{eq10}) may be explicitly expressed in terms of the complex roots $z_1$, $z_2$, and $z_3$ of the characteristic equation 
\begin{equation}
z^3=2z^2-z+1
,
\label{12}
\end{equation} 
giving 
\begin{equation}
  \tilde{d}(\ell) = C_1 {z_1}^\ell + C_2 {z_2}^\ell  + C_3 {z_3}^\ell
  ,
\label{eq20}
\end{equation}
 where 
\begin{eqnarray} 
C_1&=&(z_1 - 1)/[(z_1 - z_2)(z_1 - z_3)],
\nonumber
\\
C_2&=&(z_2 - 1)/[(z_2 - z_1)(z_2 - z_3)],
\nonumber
\\
C_3&=&(z_3 - 1)/[(z_3 - z_1)(z_3 - z_2)]
.
\label{21}
\end{eqnarray}
One of the roots of Eq.~(\ref{12}), $z_1$, say, is real, 
\begin{equation}
z_1\equiv z_d =1.75488...\, 
.
\label{22}
\end{equation} 
It determines the large $\ell$ asymptotics of the prime degeneracies $\tilde{d}(\ell)$:  
\begin{equation}
\tilde{d}(\ell) \cong \frac{z_d}{z_d^4-2}\,z_d^\ell
\label{23}
\end{equation} 
where we used the identity $C_1 = z_d/(z_d^4-2)$. The two other roots are complex conjugate numbers, 
\begin{equation}
z_{2,3} = 0.122561... \pm 0.744862...\,i
.  
\label{24}
\end{equation} 

The special case of the periodic output of length $n$ with all digits 0 has to be considered separately. 
First let us take an arbitrary digit of the input. The number of input configurations where this digit is 0 and the resulting output has only zeroes 
 is given by $\tilde{d}(n+1)$, because the fixed 0 of the periodic input plays the role of first and last digit of the configurations of a string of $n+1$ digits. 
If the chosen digit is 1, then the number of the relevant input configurations equals $1+\sum_{i=2}^{n-1} i \tilde{d}(n-i)$, where the sum over $i$ accounts for the configurations where the digit is in a group of $i$ consecutive ones, 
plus one configuration with all input digits equal to 1.
Consequently, we obtain the following expression for the degeneracy of the output with all zeroes: 
\begin{equation}
d_D(n) = 1 + \tilde{d}(n+1) + \sum_{i=2}^{n-1} i \tilde{d}(n-i)
, 
\label{eq21}
\end{equation}
which is the largest possible degeneracy of an output of a given length. Applying the recursion relation for prime degeneracies $\tilde{d}$, Eq.~(\ref{eq10}) 
 to the terms on the right-hand side of Eq.~(\ref{eq21}) 
we find that the largest degeneracy $d_D(n)$ satisfies the same difference equation as Eq.~(\ref{eq10}) 
though with different initial condition, see, e.g., Ref.~\cite{austin1978binary} and the On-line Encyclopedia of Integer Sequencies \cite{encyclopedia_integer_sequences}. We present this equation here for future reference, 
\begin{equation}
d_D(n) = 2 d_D(n-1) - d_D(n-2) + d_D(n-3)
\end{equation} 
with the initial condition  $d_D(3)=5$, $d_D(4)=10$, and $d_D(5)=17$. 
With these initial condition, we get the explicit solution of this equation, 
\begin{equation}
d_D(n) = {z_1}^\ell + {z_2}^\ell  + {z_3}^\ell
, 
\label{26}
\end{equation}
where $z_1\equiv z_d,z_2,z_3$ are given by Eqs.~(\ref{22}) and (\ref{24}), and its large $n$ asymptotics 
\begin{equation}
d_D(n) \cong z_d^n
. 
\label{27}
\end{equation}


\section{Calculating the exact degeneracy spectrum}
\label{s3p} 

\begin{figure}[t]
\begin{center}
\includegraphics[scale=0.59]{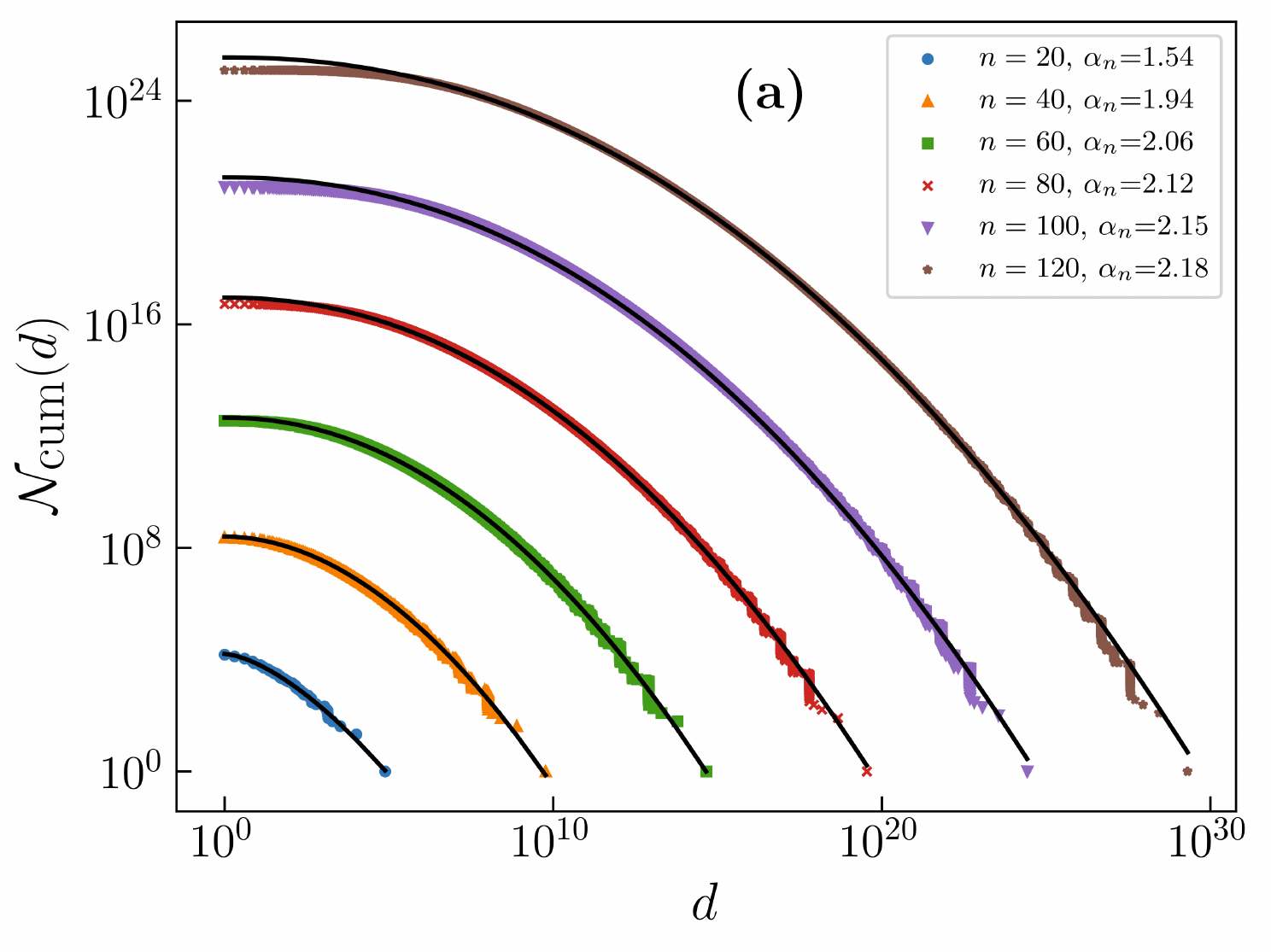}
\includegraphics[scale=0.59]{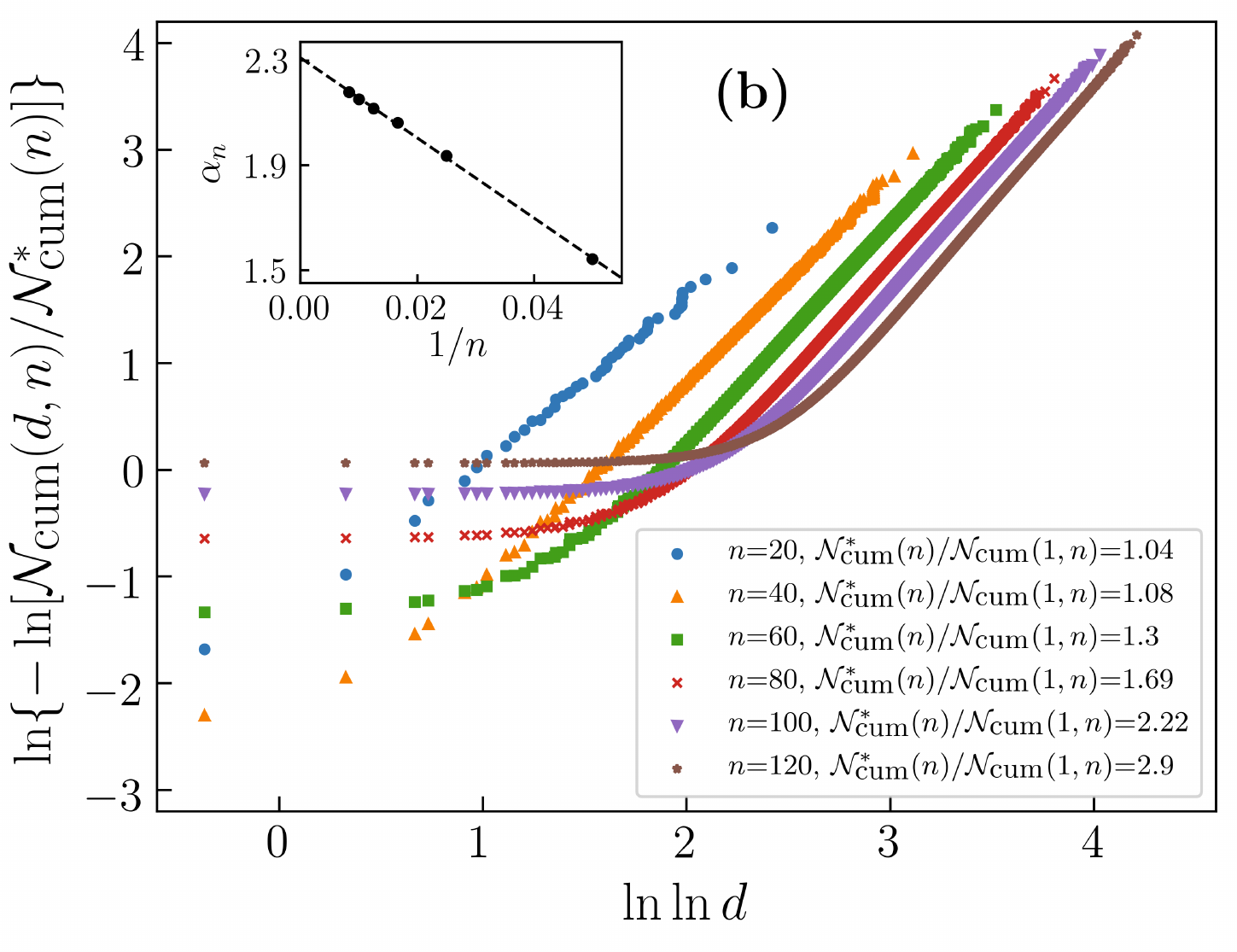}
\end{center}
\caption{ (a) Cumulative degeneracy distribution for $n=20,40,60,80,100,120$. The black curves represent least-squares fittings of $\ln{ {\cal N}_\text{cum}(d,n)}$ as $\ln{{\cal N}^\ast_\text{cum}(n)} + B_n \ln^{\alpha_n} d $ for each $n$.
(b)
Cumulative degeneracy distribution $\ln\{-\ln[{\cal N}_\text{cum}(d,n)/{\cal N}^\ast_\text{cum}(n)]\}$ vs. $\ln\ln d$ for $n=20,40,60,80,100,120$.  Inset: 
exponent $\alpha$ vs. $1/n$. 
}
\label{f4p}       
\end{figure}

In principle, for each of the $2^n$ possible inputs
 we can obtain, one by one, an output numerically by applying Eq.~(\ref{10}). 
In practice, we use a more efficient algorithm described below. 
This algorithm focuses on outputs with a fixed number of ones
 and exploits the factorization of the output degeneracies, see Eq.~(\ref{41}).  
 
The full list of degeneracies can be found from integer partitions in an explicit form, as follows.
Let us introduce the operator ${\cal P}$, which generates all integer partitions of positive integer $k$ into $r$ integers, that is, ${\cal P}(k,r)$ is the matrix whose rows $i=1,2,...,w$ are all different integer partitions $\{{\cal P}_{i1}(k,r),{\cal P}_{i2}(k,r),...,{\cal P}_{ij}(k,r),...,{\cal P}_{ir}(k,r)\}$ of $k$ into $r$ integers, $\sum_{j=1}^r {\cal P}_{ij}(k,r)=k$ \cite{andrews1998theory,andrews2004integer}. 
For an output of length $n$, we consider all partitions of $n-m$ into $m$ integers, for all possible $m=1,2,...,[n/2]$. 
For each such partition, [i.e., a row of ${\cal P}(n-m,m)$],
we find the degeneracy 
$d_i = \prod_j \tilde{d}({\cal P}_{ij}(n-m,m))$ [see Eq.~(\ref{41})]. Some of them coincide, so we find the union of them. 
Finally, we add the largest degeneracy $d_D(n)$ (corresponded to $m=0$) to the resulting set. In summary, for the full set of degeneracies ${\cal D}_\text{full}(n)$, we have
\begin{equation}
{\cal D}_\text{full}(n) =  d_D(n) \bigcup \,\Biggl\{ \bigcup_{m=1}^{[n/2]} \Bigl[ \bigcup_{i=1}^w 
\prod_{j=1}^m\tilde{d}({\cal P}_{ij}(n-m,m)) \Bigr] \!\Biggr\}
\label{29}
\end{equation}
with $ \tilde{d}(\ell)$ and $d_D(n)$ provided by Eqs.~(\ref{eq10}) and (\ref{23}), respectively. 

For each integer partition, i.e., for each row $i$ of the matrix ${\cal P}(n-m,m)$ we introduce a function giving the number 
of pieces of length $\ell$ present in this partition,
\begin{equation}\label{mu_i}
\mu_{\ell}^{(i)}(n-m,m) \equiv \sum_{j=1}^m \delta\left(\ell,{\cal P}_{ij}(n-m,m)\right),
\end{equation}
so that
\begin{eqnarray}
n-m &=& \sum_\ell \ell\mu_\ell^{(i)}(n-m,m)
,
\nonumber
\\[3pt]
m &=& \sum_\ell \mu_\ell^{(i)}(n-m,m) 
.
\label{2910}
\end{eqnarray}
One can then write 
\begin{equation}
d_i = \prod_j \tilde{d}({\cal P}_{ij}(n-m,m)) = \prod_\ell [\tilde{d}(\ell)]^{\mu_\ell^{(i)}(n-m,m)}
.    
\label{2920}
\end{equation}
The number of outputs that contain $m$ chains of zeros with lengths specified by the integer partition ${\cal P}_{i}(n{-}m,m)$ is then obtained by
considering the number of distinct permutations of the $m$ strings of zeros, multiplying by $n$, and finally dividing by $m$, giving
\begin{equation}
{\cal N}[ {\cal P}_{i}(n-m,m)] = 
n\frac{(m-1)!}{\prod_\ell \mu_\ell^{(i)}(n-m,m)!},
\label{29_1}
\end{equation}
where the product in the denominator is over the lengths of the parts of this partition. 
The total number of outputs with degeneracy $d$ is then finally:  
\begin{eqnarray}
&&
{\cal N}(d,n) =  \delta[d,d_D(n)] 
\nonumber
\\[3pt]
&&
+ \! 
\sum_{m=1}^{[n/2]}\sum_{i=1}^w n\frac{(m-1)!}{\prod_\ell \mu_\ell^{(i)}(n-m,m)!}\,\delta\Bigl[d,\!\prod_\ell [\tilde{d}(\ell)]^{\mu_\ell^{(i)}(n{-}m,m)}\Bigr]
,    
\nonumber
\\[3pt]
&&
\label{31}
\end{eqnarray}
where $\delta(a,b)$ is the Kronecker symbol. 
This is the expression we use for computing ${\cal N}(d,n)$ in an efficient way. 
For the sake of brevity, hereafter we refer to ${\cal N}(d,n)$ as a degeneracy distribution, without 
normalization. 

The distribution ${\cal N}(d,n)$ can also be built up recursively starting from small values of $d$ and $n$, as we show in Appendix \ref{a13}. 

%

\section{Output degeneracy distribution for complete input datasets}
\label{s3pp} 

Using this algorithm we obtained  the number of outputs ${\cal N}(d)$ for the full spectrum of degeneracies $d$ for $n$ up to $120$, see Supplementary Material~\cite{supplementary}.
These results demonstrate that the degeneracies $d_i$ , $i=1,...,D$, form a discrete spectrum of values where $D$ is the rank of the largest degeneracy, and $d_1=1$. 

\begin{figure}[t]
\begin{center}
\includegraphics[scale=0.59]{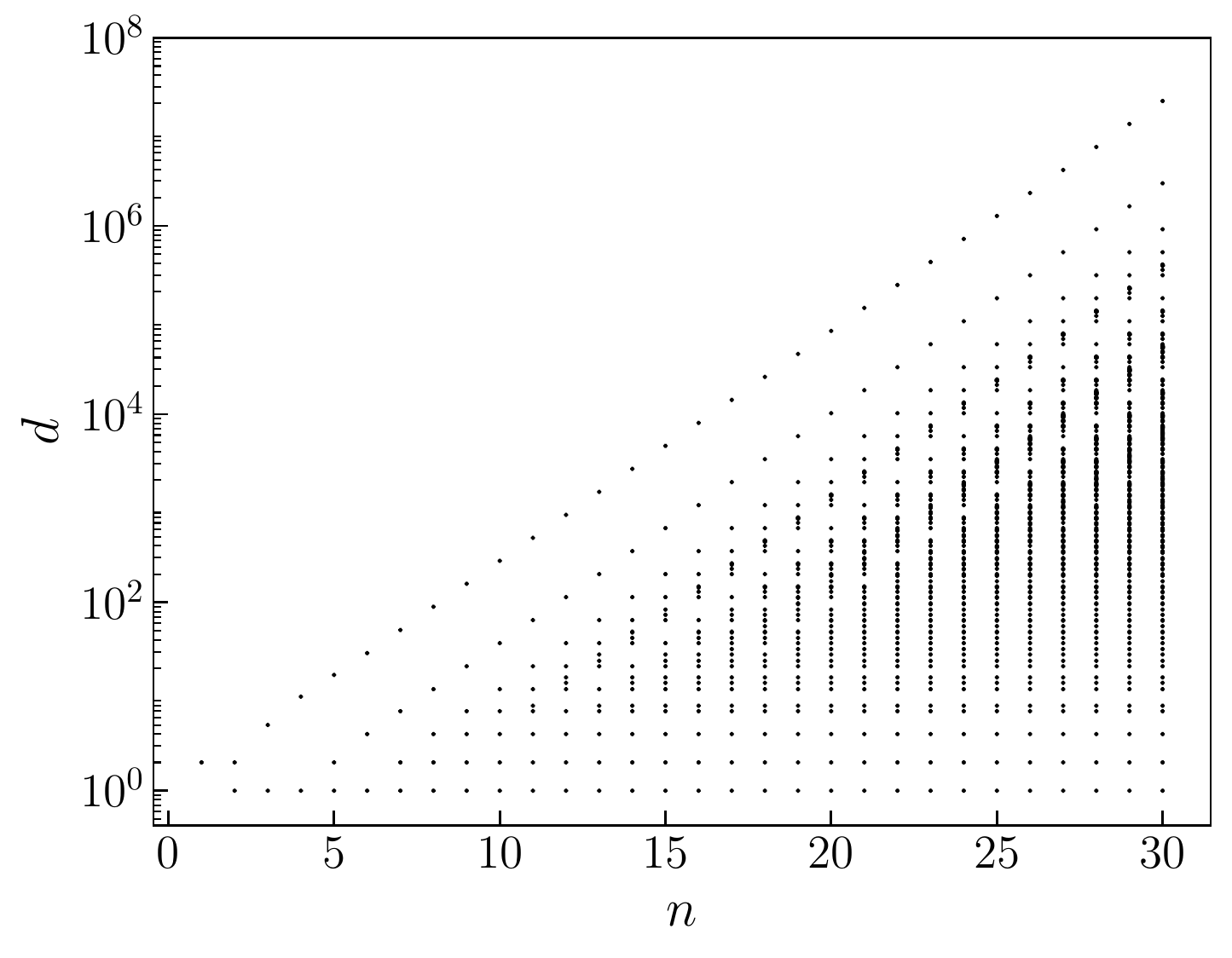}
\end{center}
\caption{
The set of degeneracy values $d$ for each $n$ (logarithmic scale). 
(The range of $n$ was selected to make individual points visible.)}
\label{f3}       
\end{figure}

Fig.~\ref{f1} shows the resulting distribution and the corresponding cumulative distribution ${\cal N}_\text{cum}(d)$ for $n=21$, $50$, and $120$. 
Here ${\cal N}_\text{cum}(d_i) \equiv \sum_{j=i}^D {\cal N}(d_j)$. In particular, ${\cal N}_\text{cum}(d_1)=M(n)$, i.e., the total number of outputs. 
Figs.~\ref{f1}(b), (c), and (d) 
demonstrate that the cumulative degeneracy distributions decay with $d$ more rapidly than a power law.  
On the other hand, the decay of the cumulative distribution is well described by the function
\begin{equation}
{\cal N}_\text{cum}(d) \propto e^{-c\ln^\alpha d} = d^{-c\ln^{\alpha-1}\! d}
,
\label{44}
\end{equation}
where $c$ is a positive number and exponent $\alpha$ 
approaches $2.3$ as $n\to\infty$, see the inset in Fig.~\ref{f4p} (b). 
Notice that the degeneracy distribution for smaller $n$, Fig. \ref{f1} (a), appears more like a power law than the degeneracy distribution for larger $n$, Fig.~\ref{f1}(c), for example, since the range of $d$ increases with $n$, while the exponent $c\ln^{\alpha-1} \! d$ of the function 
$d^{-c\ln^{\alpha-1}\! d}$ varies slowly. 
Similarly, the cumulative distribution plotted in log-log scale for $n=21$ deviates from linear (power-law behavior) noticeably less than for larger $n$. 
The wide range of degeneracies $d$ that we observe enable us to present 
Fig.~\ref{f4p} (b) showing $\ln\{-\ln[{\cal N}_\text{cum}(d)/{\cal N}_\text{cum}(1)]\}$  vs. $\ln\ln d$.  
This plot supports the functional form given in Eq.~(\ref{44}). Note that in Fig.~\ref{f4p} we assumed that the coefficient factor of the asymptotics in Eq.~(\ref{44}) is close to ${\cal N}_\text{cum}(1)$, which is justified by the results. 
Fig.~\ref{f3} shows how the set of degeneracies 
${\cal D}_\text{full}(n)$ varies with $n$, see below for more detail. 

In Sec.~\ref{s3} we derived the explicit expression for the largest degeneracy $d_D(n)$ corresponding to the output with all zeroes, 
Eq.~(\ref{26}), and found its large $n$ asymptotics, $d_D(n) \cong z_d^n$, Eq.~(\ref{27}). As is natural, ${\cal N}(d_D,n)=1$. 
The second largest degeneracy corresponds to an output with a single $1$. The third largest degeneracy is for an output with two ones separated by a single 0. The fourth largest degeneracy is of the output with two ones separated by three 0s. Clearly, ${\cal N}(d_{D-1},n)={\cal N}(d_{D-2},n)={\cal N}(d_{D-3},n)=n$. Using the asymptotics of $\tilde{d}(\ell)$, Eq.~(\ref{23}), we
find that, asymptotically, at large $n$, 
\begin{eqnarray}
&&
d_{D-1}(n) \cong \frac{z_d^n}{7.48391...} = \frac{z_d^n}{z_d^4 - 2} 
,
\nonumber
\\[5pt]
&&
d_{D-2}(n) \cong \frac{z_d^{n-2}}{z_d^4 - 2}
,
\nonumber
\\[5pt]
&&
d_{D-3}(n) \cong \frac{z_d^{n-3}}{z_d^4 - 2}
. 
\label{75}
\end{eqnarray}

One may also notice a complex structure in the cumulative distributions
Figs.~\ref{f1}(b), (d), and (f), ${\cal N}_\text{cum}(d)$ resembling a staircase, with steep jumps between steps.
The heights of these jumps are especially large in the region of high degeneracies. 
Similar structures may be observed in real systems, see for example Fig.~3 of Ref. \cite{marsili2013sampling}.
Inserting the asymptotics of $\tilde{d}(\ell)$, Eq.~(\ref{23}), into the expression for the degeneracy, Eq.~(\ref{41}), we see that outputs with the same number of ones have degeneracies exponentially close to $z_d^n/(z_d^4 - 2)^2$ if these ones are separated by many zeroes  and $n$ is large. 
The slight deviations from this asymptotic value mean that the degeneracies are split among many points falling in a narrow range.
For example, in the rightmost step, corresponding to outputs with exactly two ones, 
the number of these outputs is approximately $n^2/2$ because there are relatively few outputs in which the two ones are close together. 
This is the height of this jump.
These degeneracies are split into a set of about $n/2$ distinct values, 
corresponding to different intervals between the ones.
each one occurs in $n$ outputs corresponding to the location of the same structure in different parts of the ring. Other jumps are produced by outputs with $m$ strongly separated ones, 
or by outputs with, e.g., a pair of ones separated by one 0 with $m-2$ ones
  far from each other and from that pair, and so on. 
  This forms  the rich staircase-like structure that we observe in Figs.~\ref{f1}(b), (d), and (e) 
  and that is also reflected in Fig.~\ref{f3}.  
  Note that any {short} filter of our sort
  will produce a similar complex structure in the degeneracy distribution, as the degeneracy associated to a chain of zeroes (in output) of a given length must asymptotically grow exponentially with the length of the chain, as in Eq. (\ref{27}). 

We list the key numbers defining the asymptotics of a few example cases in Table \ref{su40}, including an example of a two dimensional input, and give the full degeneracy distributions for selected cases in Fig. \ref{fig3}. 
{
It is worth discussing the meaning of these results, and their interpretation in terms of sampling of complex systems.
In particular let us consider the results for the family of filters composed of a string of ones with a zero at each end, $010, 0110, 01110, $ etc., indicated in boldface in the Table. We can consider the length of these filters as a crude control parameter of our sampling. Intuitively, we expect shorter filters to be more informative. 
Resolution, defined as the entropy of a sample:
\begin{equation}
\label{resolution}
H[y] 
= -\frac{1}{N}\sum_{i=1}^N\log\left(\frac{d_i}{N} \right).
= -\sum_d \frac{d{\cal N}(d,n)}{N}\log\left(\frac{d}{N} \right)
\end{equation}
is a measure of the ability to distinguish, at the output, between different input states.
If the sample contains $M$ distinct outputs, all with the same degeneracy, the resolution is simply $\ln M$, see Eq. (29).
Notice that this is the case when all outputs are different (all degeneracies equal to $1$, $M=N$), and also when all outputs are the same (degeneracy equal to $N$, $M=1$).
The inverse of the probability of a particular output gives the expected number of distinct outputs, if we assume the probabilities to be uniform.
In this sense, we can regard the resolution as the expectation of the logarithm of the number of distinct states in a sample, based on the observation of one random state.
Shorter filter patterns correspond to higher resolution. 
The resolution attains its maximum value for the filter pattern $0$ or $1$, either of which results in a unique output for each input 
(in the case of $1$,  the output is identical to the input).
However in this case all outputs are distinct, and so these filters are
not informative about the system being sampled.
As shown in \cite{cubero2018minimally}, the measure which indicates the informativeness of a sample, is the entropy of the degeneracy distribution,
\begin{equation}
\label{entropy}
H[d] = -\sum_d \frac{d{\cal N}(d,n)}{N}\log\left(\frac{d{\cal N}(d,n)}{N} \right),
\end{equation} 
where $N= 2^n$ for the complete input set.
This is called the relevance.
Comparing Eqs. (29) and (30) shows that the resolution is never smaller than the relevance, $H[y] \geq H[d]$. 
As can be seen in the Table, the relevance is greater for shorter filters, but is actually zero for the shortest possible filters $0$ and $1$. (In this extreme, $z_d = 1$, while $z_g = z_a = 2$.) We measured the relevance for a variety of short filter patterns. Specifically, we find that the entropy $H[d]$ is maximal for the filter pattern extracting single ones, on which we focus in this work, and so
this pattern provides the most informative representations of the inputs

The number $z_d$ gives the asymptotics of the largest degeneracies. It quickly approaches $2$ as the filter pattern length increases. The largest degeneracies $\cong z_d^n$. Since $N = 2^n$, this means that almost all outputs concentrate in a few outputs, and in the limit, in a single state, i.e. all outputs 
are the same 
and the filter patterns are not informative. For the shortest few patterns, the value of $z_d$ quickly falls, accompanied by a rapidly increasing relevance, indicating a transition to informative sampling.
On the contrary, $z_g$, which gives the total number of outputs $M(n)$, increases with decreasing filter length, as shorter filters have more possible outputs. 
%
 Taken together, these results indicate that the maximally informative sampling for a given family of filters is the shortest pattern having length greater than $1$. This behavior is analogous to the transition observed in more complex problems (see for example \cite{cubero2018minimum}), reinforcing the relevance of our tractable model to the study of more complex problems.}

%

%
%
Importantly, the sets of outputs obtained with the larger filters are subsets of the set of the outputs obtained with our reference filter pattern $010$. For this pattern, the value $z_g$, {which gives the asymptotics of the number of different outputs,} is maximal, and the number of different outputs is the largest one. (Notice that the zero-pattern in Table~\ref{su40} has the same $z_g$ as the reference pattern but a lower entropy $H[d]$.) 
Some filters have the property that two occurrences of the pattern cannot partially overlap. All such filter patterns of a given length give an identical degeneracy distribution, as we show in the table. We suggest that the filtering problem with such non-overlapping filter patterns of an arbitrary finite length can be solved in a way similar to our reference pattern.

\begin{table*}
\centering
\begin{ruledtabular}
\begin{tabular}{p{0.1\textwidth} p{0.075\textwidth} p{0.025\textwidth} c c c c c c p{0.1\textwidth}}
& pattern && $z_g$ & $z_d$ & $z_a$ & $\!\!H[d], n{=}36$ & $\!\!H[y], n{=}36$ & $D, n{=}36$ &
\\ 
\\ 
\hline 
\\[-0.3ex]
& \textbf{010} &	&	1.61803	&  	1.75488	&	1.46557 &   6.20926 & 12.66744 &  777\\[0.7ex]
& 0\spc{-0.223}0\spc{-0.223}0 &	&   1.61803 &   1.83929 & 	1.49710 & 	5.230(5) & 10.83768 & 554(2)\\[1.5ex]        
%
\\[-0.8ex]
& \textbf{0110}   &   \multirow{2}{*}{$\left.\rule{0pt}{2.2ex}\right\}$} &   \multirow{2}{*}{1.46557} &   \multirow{2}{*}{1.86676} &   \multirow{2}{*}{1.22074} &   \multirow{2}{*}{4.27711} &   \multirow{2}{*}{8.05933} &   \multirow{2}{*}{698}\\
& 0\spc{-0.234}1\spc{-0.234}0\spc{-0.234}0\\[0.7ex]
& 0\spc{-0.234}0\spc{-0.234}0\spc{-0.234}0 &  & 1.52895 &         1.92756 &      1.41963(2) &   3.18801 & 6.36220 & 311\\[1.5ex]
%
\\[-0.8ex]
& 0\spc{-0.240}0\spc{-0.240}1\spc{-0.240}0\spc{-0.240}0 &&      1.46557&    1.9417 &  1.38028 & 2.31633  & 4.81344   &   291\\[0.7ex]
& \textbf{01110}   &   \multirow{3}{*}{$\left.\rule{0pt}{3.7ex}\right\}$}	&   \multirow{3}{*}{1.38028}	&   \multirow{3}{*}{1.93318}	&   \multirow{3}{*}{1.16730}	&   \multirow{3}{*}{2.27238} &   \multirow{3}{*}{4.88224}	&   \multirow{3}{*}{255}  \\
& 0\spc{-0.240}1\spc{-0.240}1\spc{-0.240}0\spc{-0.240}0\\
& 0\spc{-0.240}1\spc{-0.240}0\spc{-0.240}0\spc{-0.240}0\\[0.7ex]
& 0\spc{-0.240}1\spc{-0.240}0\spc{-0.240}1\spc{-0.240}0 &&      1.44327  &  1.94789 &1.32472 & 2.20202& 4.53035 &   301\\[0.7ex]
& 0\spc{-0.240}0\spc{-0.240}0\spc{-0.240}0\spc{-0.240}0 &&     1.46557(2) & 1.96595 & 1.3652(2) & 1.83904 & 3.61870 &   190\\[1.5ex]
%
\\[-0.8ex]
& 0\spc{-0.243}0\spc{-0.243}1\spc{-0.243}1\spc{-0.243}0\spc{-0.243}0 &   \multirow{2}{*}{$\left.\rule{0pt}{2.2ex}\right\}$} &      \multirow{2}{*}{1.38028}  &     \multirow{2}{*}{1.96931} & \multirow{2}{*}{1.2499(2)}  & \multirow{2}{*}{1.29800} & \multirow{2}{*}{2.84346} & \multirow{2}{*}{197} \\
& 0\spc{-0.243}0\spc{-0.243}1\spc{-0.243}0\spc{-0.243}0\spc{-0.243}0\\[0.7ex]
& 0\spc{-0.243}1\spc{-0.243}0\spc{-0.243}0\spc{-0.243}1\spc{-0.243}0 &&      1.37108(1)    &      1.97113 & 1.1938(5) & 1.29102  & 2.76917 &      218\\[0.7ex]
& \textbf{011110}  &  	\multirow{6}{*}{$\left.\rule{0pt}{8.2ex}\right\}$} &   \multirow{6}{*}{ 1.32472} &   \multirow{6}{*}{1.96717}	&   \multirow{6}{*}{1.13472}    &   \multirow{6}{*}{1.24124} &   \multirow{6}{*}{2.85788}  &   \multirow{6}{*}{123}\\
& 0\spc{-0.243}1\spc{-0.243}1\spc{-0.243}1\spc{-0.243}0\spc{-0.243}0\\
& 0\spc{-0.243}1\spc{-0.243}1\spc{-0.243}0\spc{-0.243}1\spc{-0.243}0\\
& 0\spc{-0.243}1\spc{-0.243}1\spc{-0.243}0\spc{-0.243}0\spc{-0.243}0\\
& 0\spc{-0.243}1\spc{-0.243}0\spc{-0.243}1\spc{-0.243}0\spc{-0.243}0\\
& 0\spc{-0.243}1\spc{-0.243}0\spc{-0.243}0\spc{-0.243}0\spc{-0.243}0\\[0.7ex]
& 0\spc{-0.243}0\spc{-0.243}0\spc{-0.243}0\spc{-0.243}0\spc{-0.243}0 &&      1.4176(2)	&       1.98358 & 1.32486 & 1.06843 & 2.01805 &         123\\[1.5ex]
%
\\[-0.8ex]
& 0\spc{-0.247}1\spc{-0.247}1\spc{-0.247}0\spc{-0.247}1\spc{-0.247}1\spc{-0.247}0 &&  	1.32472	&   1.98574	&   1.158(2)    &   0.75026 & 1.56692  & 129\\[0.7ex]
& \textbf{0111110} &\multirow{2}{*}{$\left.\rule{0pt}{2.2ex}\right\}$} &    \multirow{2}{*}{1.28520} &   \multirow{2}{*}{1.98386} &     \multirow{2}{*}{1.11278}  &  \multirow{2}{*}{0.72561} &  \multirow{2}{*}{1.63275} &  \multirow{2}{*}{64}\\
& 0\spc{-0.247}1\spc{-0.247}1\spc{-0.247}1\spc{-0.247}0\spc{-0.247}1\spc{-0.247}0\\[1.5ex]
%
\\[-0.8ex]
& \textbf{01111110} &&   1.25542 &    1.99203 &     1.09698  &  0.43683 &  0.91660 &  36\\[0.7ex]
& \textbf{011111110} &&  1.23205 &    1.99605 &     1.08507  &  0.26154 &  0.50786 &  25\\[0.7ex]
& \textbf{0111111110} && 1.21315 &    1.99803 &     1.07577  &  0.15385 &  0.27853 & 16\\[1.5ex]
\\[-0.8ex]
 & \parbox[c]{1em}{
      \includegraphics[scale=0.07]{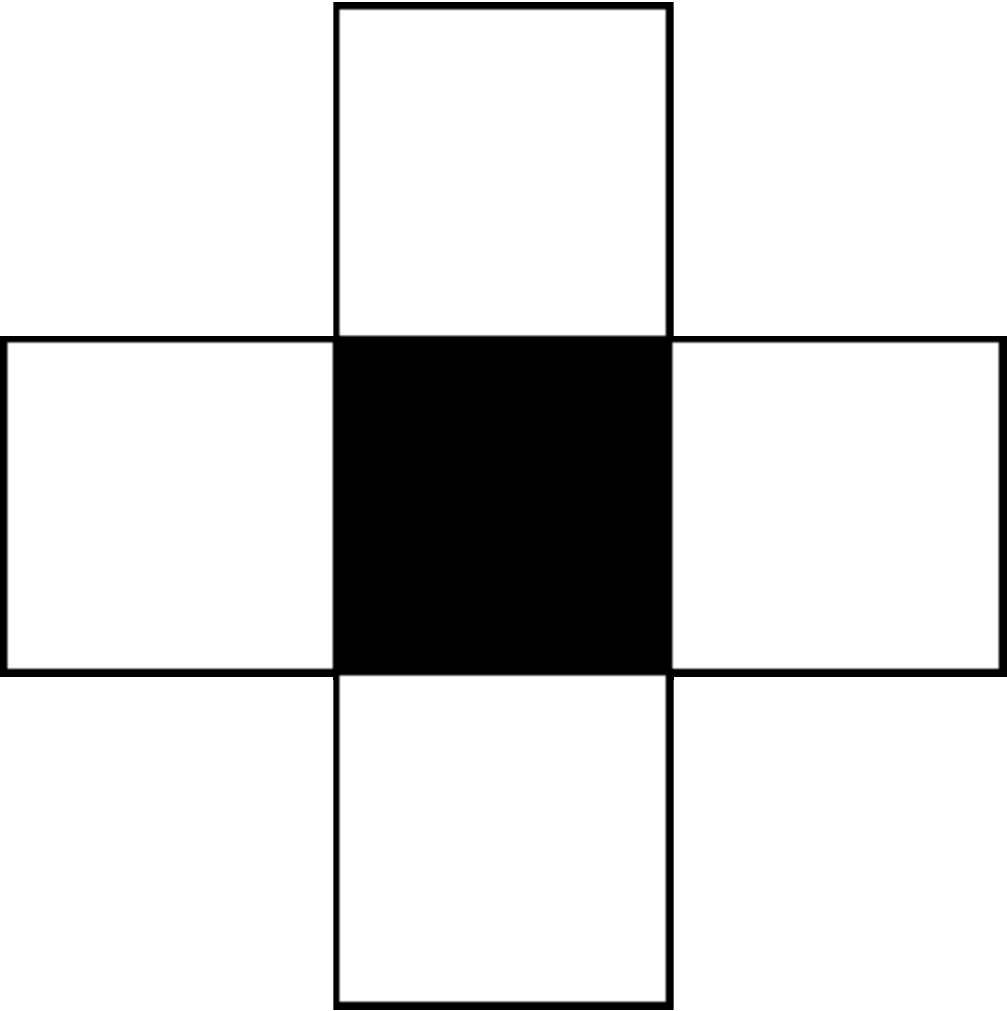}
      }
&& $1.50305$ 	& $1.9476(1)$	& 1.42(2)	& 1.46418 & 4.72664 &     624
\end{tabular}
\end{ruledtabular}
\caption{
Values of the numbers $z_g$, $z_d$, and $z_a$ for different filters. These give the asymptotics of, respectively, the total number of outputs $M(n) \cong z_g^n$, the largest degeneracy $d_D(n) \cong z_d^n$, and the number of outputs with degeneracy one, $N(1,n) \cong z_a^n$ for large $n$. 
The final row corresponds to a simple filter, as illustrated, applied to 2D input (see also Fig. \ref{fig3}). 
Note that we also included filter patterns consisting of all zeroes. 
For each filter we also give the relevance $H[d]$ (in nats) calculated from the degeneracy distribution and the resolution $H[y]$ for inputs of size $n=36$. For the sake of comparison, the standard entropy of the inputs of this size is $H=36\ln 2 = 24.95330$. Finally we include the number of distinct degeneracies $D$ for each pattern.
} \label{su40}
\end{table*}

In Appendix~\ref{a13} we derive the chain of linear coupled recursion relations 
for the number of outputs of degeneracy $d$, ${\cal N}(d,n)$, see Eqs.~(\ref{a210}) and (\ref{a220}). 
These recursions generate exact ${\cal N}(d,n)$ for finite $d$ and $n$, and in this sense provide the exact full spectrum of degeneracies. (There recursions also provide us with the explicit leading large $n$ asymptotics of ${\cal N}(d,n)$, Eq.~(\ref{a230}), for an arbitrary $d$.)
In particular, Eq.~(\ref{eq90}) with the initial conditions, Eq.~(\ref{eq105}), provides ${\cal N}(1,n)$ which is the number of configurations having only groups of zeroes of length 1, 2, and 3 between single ones. 
We find the large $n$ asymptotics for this number:  
\begin{equation}
{\cal N}(1,n) \cong z_a^n
,  
\label{100}
\end{equation}
where 
\begin{equation}
z_a =1.46557...
\label{101}
\end{equation}
is the real root of the characteristic equation $z^4 = z^2 + z + 1$.

Note that there are three key constants in this problem, $z_g$, $z_d$, and $z_a$, which appear in the main asymptotics: 
$M(n) \cong z_g^n$, $d_D(n) \cong z_d^n$, and ${\cal N}(1,n) \cong z_a^n$. 
Indeed, $z_g$ enters the distribution ${\cal N}(d,n)/M(n)$ after normalization, $z_d$ enters the asymptotics for degeneracies, see, e.g., Eq.~(\ref{23}) for $\tilde{d}(\ell)$ and (\ref{75}) for $d_i(n)$, and $z_a$ enters the asymptotics for ${\cal N}(d,n)$, Eqs.~(\ref{eq117}) and (\ref{a230}). 
In fact, all moments of ${\cal N}(d,n)$ exponentially diverge with $n$, $\sum_{i=1}^D d_i^k {\cal N}(d_i,n) \cong c_k^n$, where the numbers $c_k$ and their large $k$ asymptotics are presented in Appendix~\ref{a3}. Clearly, for the first moment we have $\sum_{i=1}^D d_i\, {\cal N}(d_i,n) = 2^n$.

\begin{figure}[t]
\begin{center}
\includegraphics[scale=0.59]{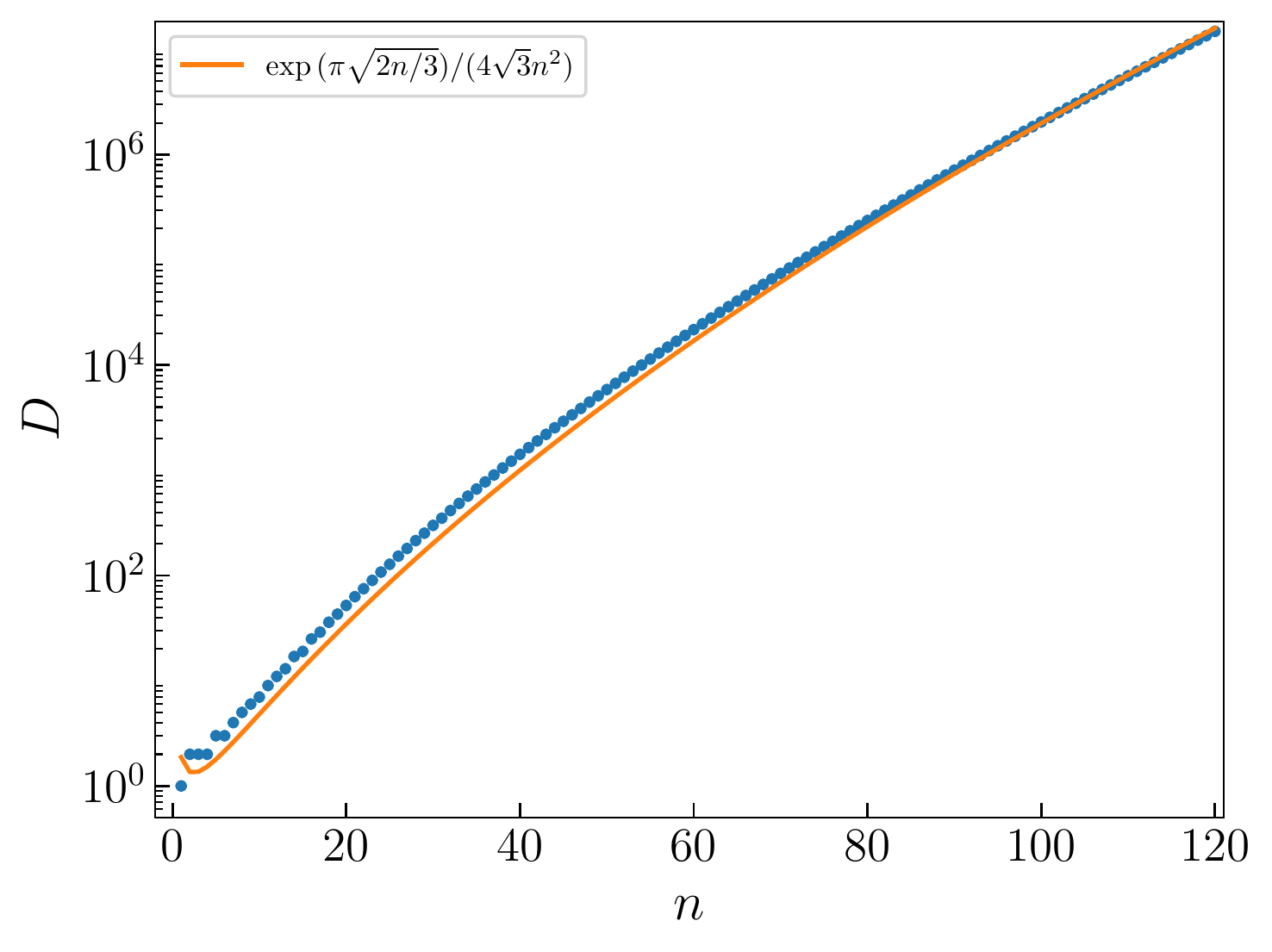}
\end{center}
\caption{
Total number of different degeneracies $D$ vs. $n$ (symbols). The curve is the number of integer partitions of $n$ divided by $n$.
}
\label{f4}       
\end{figure}

Using Eq.~(\ref{29}) we obtain the full set of degeneracies
${\cal D}_\text{full}(n)$ for different $n$, shown in Fig.~\ref{f3}, and the series of total number of discrete values of degeneracy $D(n)$ versus input size, represented in Fig.~\ref{f4}, see also the Supplementary Material~\cite{supplementary}.
 As is natural, $D(n)$ is smaller than the number $p(n)$ of integer partitions of $n$. Fig.~\ref{f4} demonstrates that $D(n)$ is close to $p(n)/n$ for large $n$. The well known large $n$ asymptotics $p(n) \cong \frac{1}{4\sqrt{3}n}  e^{\pi\sqrt{2n/3}}$ \cite{hardy1918asymptotic,uspensky1920asymptotic,wilf2000lectures} enables us to estimate $D(n)\sim e^{\pi\sqrt{2n/3}}$. 

\begin{figure}[t]
\begin{center}
\includegraphics[scale=0.58]{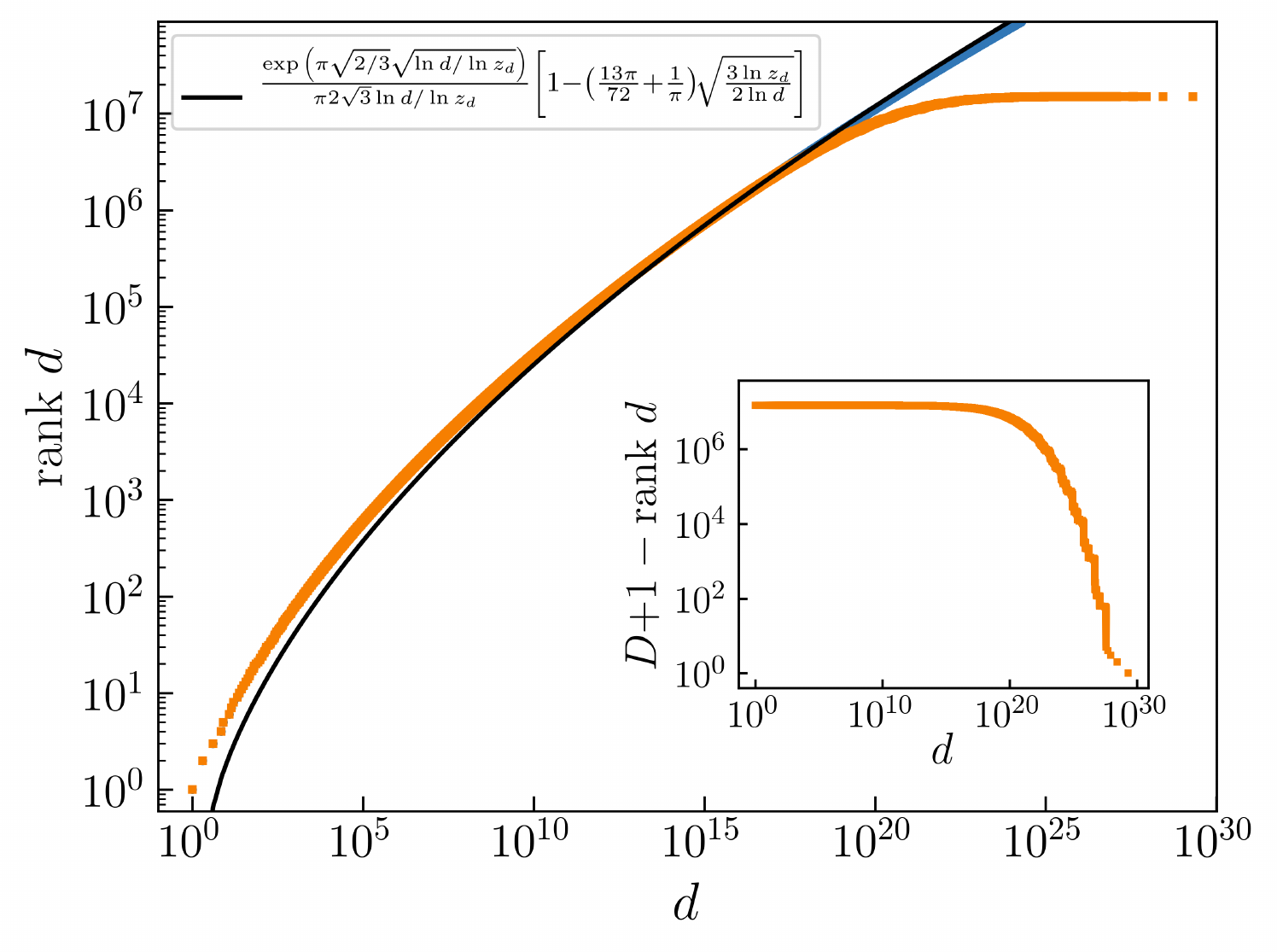}
\end{center}
\caption{
Number of distinct degeneracies less than or equal to $d$ vs. $d$. 
Inset: number of different degeneracies higher than or equal $d$ vs. $d$. 
The orange square symbols are full results for $n=120$. The circular blue symbols are values calculated in the limit $n\to\infty$ as described in Sect.~\ref{s3pp}. Solid black curve is the asymptotic approximation Eq.~(\ref{102}).
}
\label{f5}       
\end{figure}

By ranking the full set of degeneracies ${\cal D}_\text{full}(n)$ for $n=120$ we arrive at Fig.~\ref{f5}, where the main plot presents the number of different degeneracies lower than or equal to $d$ vs. $d$, increasing roughly as $\exp[\pi\sqrt{2\ln d/3\ln z_d}]$ in the region $1 \ll d \ll d_D$, and the inset shows the number of different degeneracies higher than or equal $d$ vs. $d$. The latter demonstrates a staircase-like structure similar to that of ${\cal N}_\text{cum}(d)$.

In Fig.~\ref{f5} we also plot the number of different degeneracies lower than or equal to $d$ in the infinite system, i.e., the rank of each degeneracy $d$. To obtain the full set of different degeneracies for $n\to \infty$ up to some $d_\textrm{max}$ we generate the set of prime degeneracies $\tilde{d} \leq d_\textrm{max}$. We start by listing all powers $m\geq 1 $ of the first prime degeneracy $\tilde{d}$ while $\tilde{d}^m\leq d_\textrm{max}$. Then, we multiply each member of that list by increasing powers of the next prime degeneracy, while the product stays lower or equal to $d_\textrm{max}$, and so on with all the remaining prime degeneracies $\tilde {d}\leq d_\textrm{max}$.
This procedure will result in duplicate degeneracies that should be removed, particularly for those values of $d$ that have non-unique multiplicative partitions in terms of the prime degeneracies. 

For example, the two smallest (larger than $1$) prime degeneracies are $\tilde{d}(4) = 2$ and $ \tilde{d}(5) = 4$, so a multiplicative partition of $d$ with at least a part equal to $ \tilde{d}(5) = 4$ is not unique because there is at least another partition of $d$ where the contribution of the $4$ for given in terms of $\tilde{d}(4) $ as $2^2$.
However, apart from the partitions with parts equal to $4$, the non-unique partitions are very rare, see Appendix~\ref{a13}.
In fact $\tilde{d}(4) = 2$ is the only prime degeneracy $\tilde{d}(\ell)$ that can be expressed as a product of lower $\tilde{d}$ for all $\ell\leq 5000$. 

The rank of the degeneracies in the infinite system can be estimated with precision for large $d$, as shown by the black line 
in Fig.~\ref{f5}.
For large $\ell$ the logarithms of prime degeneracies are uniformly distributed $\ln \tilde{d}(\ell) \approx \ell \ln z_d$. 
Assuming that all degeneracies $d$ have a unique factorization in terms of the prime degeneracies $\tilde{d}$,
the expected number of different degeneracies smaller than $d \sim z_d^m$ would be $\textrm{rank\ } d \approx \sum_{n \leq m} p(n)$, where $p(n)$ is the number of integer partitions of $n$.
The vast majority of the degeneracies for which the assumption does not hold are the values of $d$ which are multiples of $4$, because that factor $4$ can also be expressed as $2^2$.
We therefore remove from each term of the previous sum the number of integer partitions that have at least one factor equal to $2$, which is given by $p(n-2)$, and write for large $d$:
\begin{eqnarray}
&& \textrm{rank } d  \approx \sum_{n \leq \ln d / \ln z_d} [ p(n) - p(n-2) ] 
\nonumber
\\
&&\approx p(\ln d / \ln z_d) + p( \ln d / \ln z_d-1) 
\nonumber
\\
&&\approx \frac{\exp\left( \pi \sqrt{2/3} \sqrt{\ln d /\ln z_d} \right)}{\pi 2 \sqrt{3} (\ln d /\ln z_d) } \left[1 {-} \left(\frac{13 \pi}{72} {+} \frac{1}{\pi}  \right)\sqrt{\frac{3  \ln z_d }{ 2 \ln d }}  \right],
\nonumber
\\
\label{102}
\end{eqnarray}
where we have used the two leading terms of the asymptotic expansion of the number of integer partitions~\cite{hardy1918asymptotic}
\begin{equation}
p(n) \approx \frac{\exp\left( \pi \sqrt{2/3} \sqrt{n} \right)}{4 \sqrt{3} n } \left[1 {-} \left(\frac{\pi}{72} {+} \frac{1}{ \pi}  \right)\sqrt{\frac{3}{ 2 n}}  \right].\label{p_n}
\end{equation}
Fig.~\ref{f5} demonstrates that $\textrm{rank } d$ as a function of $d$ is well described by the estimate Eq.~(\ref{102}).

\section{Mean-field theory}
\label{s3ppp}  

Here we develop a mean-field theory enabling us to describe 
the cumulative distribution of degeneracies, ${\cal N}_\text{cum}(d)$, in the region of large $d$ where there are few ones 
 in the outputs, and so the gaps of zeroes 
 between them are typically large. In this situation one can assume that the  ones 
 exist in a sea (or a mean field) of zeroes,  
far from each other, so that the degeneracy of an output is completely determined by the number of its ones 
 (and the output size $n$). 

This ansatz is based on the observation that the three terms on the right-hand side of Eq.~(\ref{eq20}) behave very differently for increasing $\ell$:
The first term grows exponentially as $z_d^\ell$, where $z_d\equiv z_1=1.754877...$ is real.
The combined contribution of the other two terms is also real, since $z_2=z_3^*$ (here $^*$ denotes the complex conjugate). It decays exponentially because $\left| z_2 \right| = \left| z_3 \right| = 0.754877...<1$.
Thus, for increasing $\ell$, the deviation of the asymptotics  
\begin{equation}
  \tilde{d}(\ell) \cong C_1 {z_d}^\ell
\label{eq30}
\end{equation}
from the exact value of $\tilde{d}(\ell)$ approaches $0$ exponentially rapidly, making this approximation 
excellent for large $\ell$.
Consider an output with $m$ ones,
separated by $m$ strings of zeroes 
 of length $\ell_1$, $\ell_2$, ..., $\ell_k$. 
If all $\ell_i \gg 1 $ the degeneracy $\overline{d}(m)$ of this output with $m$ ones 
 is accurately given by the asymptotic expression 
\begin{equation}
\overline{d}(m) = \prod_{i=1}^{m} C_1 {z_d}^{\ell_i} =  {C_1}^m  {z_d}^{n-m}
,
\label{eq40}
\end{equation}
where we used the 
condition $\sum_{i=1}^m \ell_i = n-m$ that 
the number of ones plus the number of zeroes must be equal to the total number of digits $n$. 

The total number of different outputs with $m$ ones is
\begin{equation}
\overline{\cal N}(m) 
= \frac{n}{n-m} {{n-m}\choose{m}}
,
\label{eq50}
\end{equation}
see Eq.~(\ref{eqa5}) in Appendix~\ref{a11}. 
Notice that $\overline{\cal N}(m)$ 
coincides with the $m$-th term of the sum in Eq.~(\ref{20}), as it must.

One should stress that although we derived the 
expressions for $\overline{d}(m)$ 
and $\overline{\cal N}(m)$ 
for $m>0$, these expressions are equally valid in the special case of $m=0$. 
Indeed, Eq.~(\ref{eq50}) gives $\overline{\cal N}(0)=1$, which is exact. 
Further, Eq.~(\ref{eq40}) gives $\overline{d}(0)\cong z_d^{n}$ for large $n$ coinciding with $d_D\cong {z_d}^{n}$, Eq.~(\ref{27}). 
Thus Eqs.~(\ref{eq40}) and (\ref{eq50}) 
work very well also for $m=0$. 
For $m=1$ we find a similar situation in terms of accuracy. In this case there is a single string of zeroes 
 of size $n-1$, then every output with a single 1 has degeneracy equal to $\tilde{d}(n-1)$.
The deviation between $\overline{d}(1)= C_1 z_1^{n-1}$ 
and $\tilde{d}(n-1)$ decays exponentially with $n$:
For each value of $m>1$ there is actually more than one possible value for the degeneracy, that depends on the particular distribution of lengths of the strings of zeroes. 
Nonetheless, for large $n$ and $m \ll n$ most of the outputs contain only large strings of zeroes.
Therefore the set of points ($\overline{d}, \overline{\cal N}_\text{cum})$ for $k=0,1,2,...$ describes the exact cumulative distribution ${\cal N}_\text{cum}(d)$  well at the points $d=\overline{d}(m)$ in the region of large $d$, which corresponds to $m\ll n$. There are jumps in the cumulative distribution at these points when $n$ is large, and $\overline{\cal N}_\text{cum}(m)$ approaches the top points of these jumps. 

The resulting mean-field theory expression for the distribution takes the following form:
\begin{eqnarray}
\!\!\!\!\!\!\!\!\!
\overline{{\cal N}}(d,n) & = & \delta(d,d_D(n)) 
\nonumber
\\[3pt]
& + &
\sum_{m=1}^{[n/2]} \frac{n}{n-m} {{n-m}\choose{m}}\,
\delta\Bigl(d,\frac{z_d^n}{(z_d^4-2)^m}
\Bigr)
.   
\label{5000}
\end{eqnarray}
This approximation is compared with the exact cumulative distribution in Fig.~\ref{f1}.
The approximation can also be obtained from the exact result for ${\cal N}(d,n)$, Eq.~(\ref{31}), by approximating the weighted sum $\sum_i$ of Kronecker symbols in the expression for ${\cal N}(d,n)$ by a single Kronecker symbol with a factor. 
One may also obtain the asymptotics of this distribution, as shown in Appendix \ref{a5}.

\section{Degeneracy distributions of outputs for randomly generated input datasets}
\label{s4}  

\begin{figure*}[t]
\begin{center}
\includegraphics[scale=0.45]{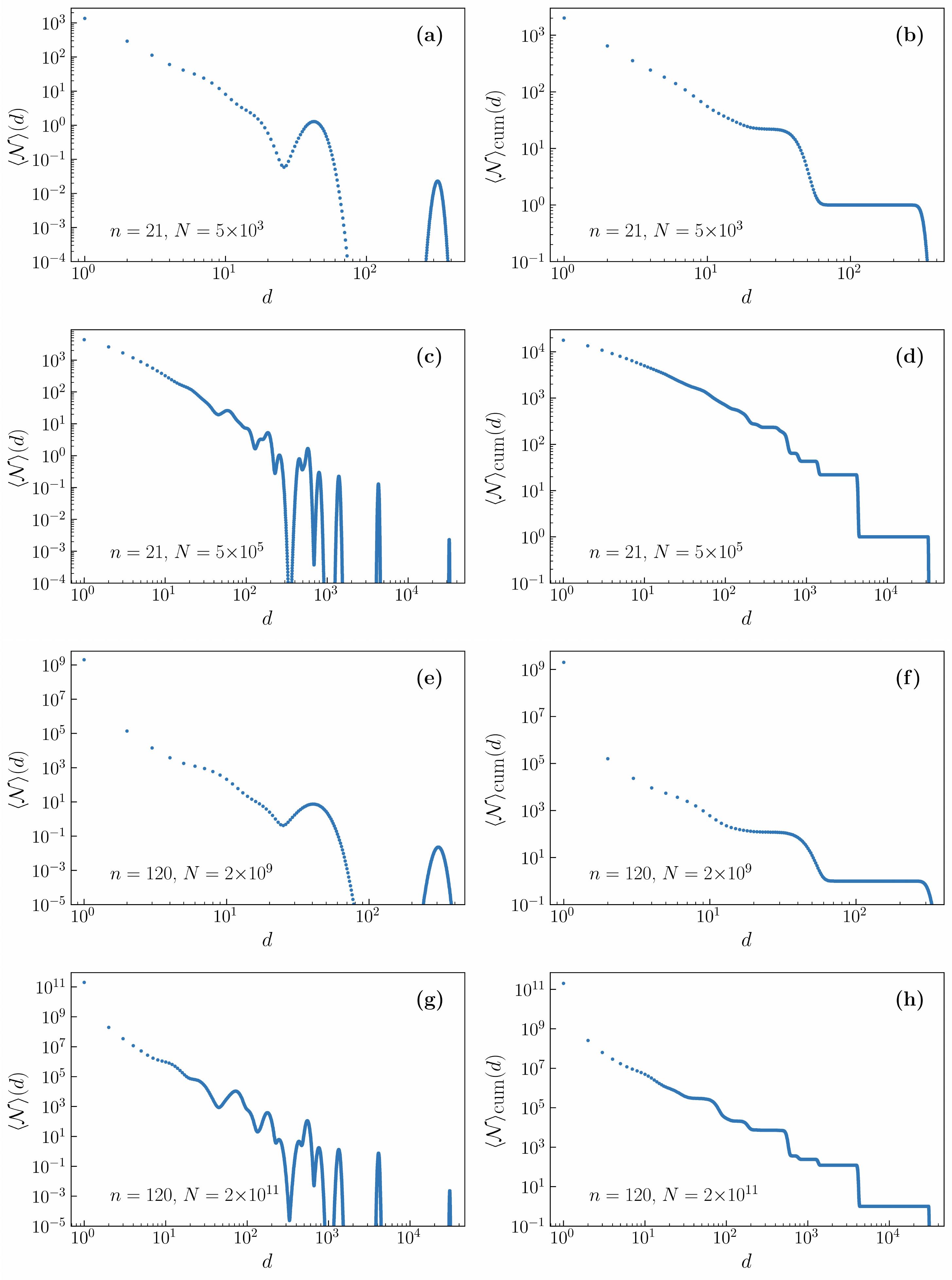}
\end{center}
\caption{
Double logarithmic scale plots of (a,c,e,g) the degeneracy distributions and 
(b,d,f,h) the cumulative degeneracy distributions obtained for a randomly generated input data sets of different sizes  $N$ for $n=21$ and $n=120$. The specific sizes of input data sets for $n=120$ are chosen to produce distributions similar to those for $n=21$. 
}
\label{f11}       
\end{figure*}

We observed in the previous section that the outputs generated by complete input data sets do not produce real power laws. 
One should note, however, that in empirical studies, input data sets, typically, are not complete. Usually, when inputs (input size $n$) are sufficiently large, the input data set sizes, $N$, are much smaller than of complete data sets, $N \ll C^n$, $C>1$. Based on the sets of outputs of complete data sets, which we obtained in the previous section (listed in the Supplementary Material \cite{supplementary}), here we find and explore the distribution of outputs of randomly generated data sets of various sizes. (In principle, these sizes can also be bigger than or equal to that of complete data sets.)

Let the size of a randomly generated input data set be $N$, i.e., we apply filtering to $N$ randomly generated rings of $n$ zeroes and ones. 
We assume that all degeneracies of the outputs of the corresponding complete input data set are known, namely the full set of pairs $\{ d_i, {\cal N}(d_i) \}$, where $i=1,2,...,D$, $D$ is the total number of degeneracies for this $n$ in the case of a complete input data set. Then for the randomly generated input data set we obtain the following expected number of outputs of degeneracy $d$: 
\begin{equation}
\langle {\cal N}\rangle(d) = {N \choose d} \sum_{i=1}^{D} {\cal N}(d_i) \Bigl(\frac{d_i}{2^n} \Bigr)^{\!d} \Bigl(1-\frac{d_i}{2^n}\Bigr)^{\!N-d}
.  
\label{120}
\end{equation}
The probability that a randomly generated input string produces a given output is simply $d_i/2^n$, where $d_i$ is the degeneracy of the output with respect to the total input set. The probability that $d$ of the $N$ inputs produce this output is then simply given by a binomial expression. Summing over all total degeneracies (multiplying by the number of instances of a given degeneracy $d_i$ and summing over $i$) gives the above expression.
Here outputs of any degeneracy within the interval $1 \leq d \leq d_D$ are present with nonzero probability, in contrast to the case of the complete input data set. The results of the application of this formula coincide with those obtained by recording statistics of outputs directly filtered from randomly generated inputs. 

Let us apply Eq.~(\ref{120}), for instance, to the cases $n=21$ [see Fig.~\ref{f1}(a) above for the complete input data set consisting of all $2^{21}$ configurations] and $n=120$, 
and inspect the distributions of outputs of uniformly randomly generated input sets of different sizes.
The results are shown in Fig.~\ref{f11}.
For each size of the random input data set we present the degeneracy distribution and its cumulative counterpart. 
Interestingly, the form of these distributions is remeniscent of samples from real systems, see for example Fig.~5 of Ref.~\cite{marsili2013sampling}.
These figures demonstrate that for $N\ll 2^n$, the distributions indeed resemble a power law. 
The reason is that for sufficiently small $N$, the distribution does not approach the large values 
of $d$ for which the variation of the exponent $c\ln^{\alpha-1} d$ in ${\cal N}_\text{cum}(d) \propto d^{-c\ln^{\alpha-1} d}$, 
[see Eq.~(\ref{44})], becomes noticeable.
As $N$ approaches $2^n$, the distributions become closer to their counterparts for the complete input data set. Clearly, the distributions obtained in the limit $N\to\infty$ will coincide with those found for the complete input data set.

Curiously, one may obtain distributions with a very similar form for different values of $n$ by choosing $N$ in order to maintain the scaling variable
\begin{equation}\label{scaling_relation}
s = (z_d/2)^n N
\end{equation}
constant.
 For example, compare the first with the third row and the second with the fourth in Fig.~\ref{f11}. 
This combination follows from the fact that the binomial product in Eq.~(\ref{120}), as a function of $d$, forms a narrow peak centered at $d_i N/2^n$, which produces $(z_d/2)^n N$ for the largest degeneracy $d_D \cong z_d^n$.  
This scaling disappears in the region of small degeneracies. 
Note that for small $N$,
 Eq.~(\ref{120}) gives $\langle {\cal N}\rangle(1) \approx N$, since $\sum_i d_i\,{\cal N}(d_i)\cong2^n$.
 In fact this is true as long as
$N\ll (4/c_2)^n$, where $\sum_i d_i^2 {\cal N}(d_i)=c_2^n$ with $c_2=3.139...$,  see Appendix~\ref{a3}. 
Thus 
the vast 
majority of outputs have degeneracy $1$ in this situation. 


\section{Discussions and conclusions}
\label{s5}  

Our straightforward, purely combinatorial treatment reveals features of distributions of outputs hidden from other approaches.  
For complete input data sets passed through our filter, we have obtained degeneracy distributions markedly distinct from power laws. On the other hand, these distributions decay as ${\cal N}_\text{cum}(d) \propto e^{-c\ln^\alpha \!d}$, $\alpha{>}1$, much slower than exponentially, and in this sense they can still be called ``critical''. We have observed that the entire form of these output distributions essentially depends on the input size $n$, which strongly differs, for example, from heavy tailed degree distributions of complex networks having exponential cutoffs \cite{clauset2009power,dorogovtsev2003evolution}.

For randomly generated input data sets, we found degeneracy distributions which could easily be taken for power laws in empirical studies, if the data set size $N$ is essentially smaller than $2^n$. As $N\to\infty$, these distributions approach the clearly non power-law shape of the distributions for the complete input data set. 
Thus we show that the size of an input data set matters for these problems. 

Our model filter can be used as a convenient reference filtering problem. 
Although apparently rather different in construction, we argue that our problem can serve as a tractable representative model for sampling of complex systems.
We focused on simple input sets which were uniformly random strings of zeroes 
 and ones. 
Correlated inputs 
are more challenging for analytical treatment. Exploring the simplest filter patterns, we showed that the statistics of outputs is determined not by the form of filter patterns but rather by what occurs in the gaps of zeroes
 between them. 
The degeneracy corresponding to each such gap can be found using recursion relationships.
We then used an integer partitions apparatus to aggregate the statistics of prime degeneracies from these gaps, finding the exact full spectrum of output degeneracies.

Considering the permutations of the integer partitions allows us to calculate the resulting exact degeneracy distribution ${\cal N}(d,n)$. Alternatively, using multiplicative partitions we derive coupled linear recursions providing the exact ${\cal N}(d,n)$ for any finite $d$ and $n$ and the explicit large $n$ asymptotics.
Finally we developed a mean-field theory which describes the approximate degeneracy spectrum, degeneracy distribution and their asymptotic behavior. Our mean-field theory of Sec.~\ref{s3p} also derives from the gaps between outputs.

These results show that in filter problems of this kind, the statistics of outputs is determined by the gaps between outputs, which are essentially determined by the size of a filter pattern. Longer filter patterns than ours, Eq.~(\ref{10}), can be subjected to a similar analysis, see Fig. \ref{fig3} and Table \ref{su40}. 
Longer filter patterns, which give less information about the input string, indeed produce less informative degeneracy distributions (i.e. having lower entropy).

For the sake of simplicity, we studied inputs containing only zeroes 
 and ones. 
We expect that our approach and the mean-field theory should be applicable to more rich inputs containing more degrees of freedom: larger sets of numbers, vectors, etc., as well as to more general cooperative systems with a large number of local minima. 
The next natural step after the mean field theory, namely a fluctuation theory, should be based on accounting for small gaps between ones and the fluctuations of the lengths of these gaps.    

In summary, we have presented a representative filtering problem for which we can obtain exact and complete results for the degeneracy distributions. Our methods may be generalized to other filtering and compression problems involving more complex filter patterns and complex, not necessary synthetic, higher dimensional inputs. We believe the qualitative insights gained from studying this system can be applied to information processing problems in general.

\begin{acknowledgments}

This work was supported by National Funds through
FCT--Portuguese Foundation for Science and Technology,
I.P. Project No. IF/00726/2015. R.~A.~C. acknowledges the FCT grant SFRH/BPD/123077/2016. 
\end{acknowledgments}


\appendix

\section{Recursive relations}
\label{a1}

Here we present the derivations of the expression for the number of different values of degeneracy of outputs produced by complete input data sets, Eq.~(\ref{20}), and the recursions for ${\cal N}(d,n)$ used in Sec.~\ref{s3}.


\subsection{Derivation of Eq.~(\ref{20}) for $M(n)$}
\label{a11}

The total number $M(n)$ of different outputs in our problem is the number of all possible periodic (period $n$) combinations of zeroes and ones having no neighboring ones. 
Clearly, this number coincides with the total number of different combinations of dimers in a ring of length $n$. 

To find the number of different outputs with $k$ ones, that is, the number of ways of placing $k$ dimers in the ring, 
it is convenient to consider the cases when the first output digit is 1 and when it is 0 separately. 
This is equivalent to fixing the state of two chosen neighboring units: either there is a dimer connecting this pair or this dimer is absent.

(i) When the first digit of an output is 1, both the second and the last digits must be 0, and the number of outputs in this case is given by the binomial coefficient ${n-k-1}\choose{k-1}$, which corresponds to starting with the sequence $1,\{0\}_{n-k}$ (where $\{0\}_{n-m}$ is a sequence of $n-m$ zeroes), 
and choosing $k-1$ out of $n-k-1$ distinct zeroes 
 to be replaced by $01$. 
Due to the periodic boundary condition, the last 0 cannot be selected for replacement, hence the number of zeroes that can be replaced is $n-k-1$.

(ii) When the first digit of an output is 0, we start from the sequence $\{0\}_{n-k}$ and replace $k$ zeroes 
 by 01. 
In this case the replacement may be made at any of the $n-k$ zeroes, 
and the number of outputs is ${n-k}\choose{k}$. 

The total number of different outputs is obtained by summing these two contributions
\begin{equation}
\overline{\cal N}(k) 
= {{n-k-1}\choose{k-1}} + {{n-k}\choose{k}} = \frac{n}{n-k} {{n-k}\choose{k}}
, 
\label{eqa5}
\end{equation}
where we used the notation $\overline{\cal N}(k)$ from Sec.~\ref{s3ppp}.

Summing over all possible values of $k$ gives
\begin{eqnarray}
M(n) & =  & 1+ \sum_{k=1}^{[n/2]}\Bigl[ \frac{(n-k-1)!}{(k-1)!(n-2k)!} + \frac{(n-k)!}{k!(n-2k)!}  \Bigr] 
\nonumber
\\[5pt]
& = & n \sum_{k=0}^{[n/2]}\frac{1}{n - k} {n - k \choose k}
,
\label{a10}
\end{eqnarray}
i.e., Eq.~(\ref{20}). 


\subsection{Recursions for ${\cal N}(d,n)$}
\label{a13}

Chains of exactly three zeroes in the output have no degrees of freedom in the input; they can only be produced in one way. Thus outputs containing no chains of zeroes of length greater than three have degeneracy $1$.
For the input size $n$, the number of such outputs, ${\cal N}(d,n)$, can be obtained recursively. 
Here we derive the recursive relation for ${\cal N}(1,n)$, ${\cal N}(2,n)$, and ${\cal N}(4,n)$, and indicate the recursions for higher degeneracy $d$. See also Supplementary Material \cite{supplementary}.

Sequences of degeneracy 1 can be regarded as assortments of three kinds of building blocks, $01$, $001$, and $0001$, put together in a ring of length $n$. 
All configurations of blocks are allowed, as long as the total number of binary digits is $n$.
Let us consider a particular position $i$ in the ring and the block to which $i$ belongs.
If we add a block $01$ between the block of $i$ and one of its neighbor blocks (say, the one to the right) to every possible configuration of length $n-2$, we get every possible configuration of length $n$ that has a block $01$ to the right of the block of $i$. 
Doing the same with configurations of length $n-3$ and blocks $001$, we get all configurations with a block $001$ to the right of the block of $i$. 
Finally, repeating the procedure for configurations of length $n-4$ and blocks $0001$, gives all configurations with a block $0001$ to the right of the block of $i$.
Since every block must be $01$, $001$, or $0001$, the union of these three sets is the full set of configurations of degeneracy 1 in system of $n$ digits.
Thus, for the number of configurations in this set, ${\cal N}(1,n)$, we can write
\begin{equation}
{\cal N}(1,n) = {\cal N}(1,n-2) + {\cal N}(1,n-3) + {\cal N}(1,n-4)
.
\label{eq90}
\end{equation}
The explicit solution of this linear difference equation is given in terms of the roots, $z_1$, $z_2$,  $z_3$, and $z_4$, of the characteristic equation $z^4 = z^2 + z +1$:
 \begin{equation}
{\cal N}(1,n) = {z_1}^n + {z_2}^n + {z_3}^n + {z_4}^n
,
\label{eq100}
\end{equation}
where the coefficients of the powers of the roots $z_i$ are found through the initial condition, 
 \begin{equation}
{\cal N}(1,1){=}0, \  {\cal N}(1,2){=}2, \ {\cal N}(1,3){=}3, \ {\cal N}(1,4){=}6
,
\label{eq105}
\end{equation}
and are equal to 1. The root $z_1 \equiv z_a = 1.46557...$ determines the large $n$ asymptotics of ${\cal N}(1,n)$, Eq.~(\ref{100}).

We can expand the analysis to higher values of degeneracy.
For example, the second smallest value, $d_2=2$, corresponds to the outputs with blocks of the types $01$, $001$, and $0001$ and a single string $00001$.
Applying the same reasoning as above, we get three terms similar to the ones of Eq.(\ref{eq90}).
We must, additionally, consider configurations with degeneracy 1, to which we add a string $00001$ in the same way, instead of one of the other blocks.
Then we get
\begin{multline}
{\cal N}(2,n) = {\cal N}(2,n{-}2) + {\cal N}(2,n{-}3) + {\cal N}(2,n{-}4)\\ + {\cal N}(1,n{-}5)
\label{eq110}
\end{multline}
with the initial condition 
\begin{equation}
{\cal N}(2,1)={\cal N}(2,2)={\cal N}(2,3) = {\cal N}(2,4) = 0, \ \ {\cal N}(2,5)=5
.
\label{eq112}
\end{equation}
In a similar way, taking into account that the prime degeneracy $\tilde{d}(\ell)$ correspond to the inserted block of $\ell+1$ (chain of $\ell$ zeroes and 1 on the right), i.e., $2, 4, 7, 12, 21, 37, 65, ...$ correspond to $\ell+1=5,6,7,8,9,10,11,...$, respectively, we derive the recursion relations for higher ${\cal N}(d,n)$ [here $d\neq d_D(n)$, for which, clearly, ${\cal N}(d_D(n),n)=1$], 
\begin{multline}
{\cal N}(4,n) = {\cal N}(4,n{-}2) + {\cal N}(4,n{-}3) + {\cal N}(4,n{-}4) \\ + {\cal N}(2,n{-}5) + {\cal N}(1,n{-}6)
\label{eq112a}
\end{multline}
with the initial condition 
\begin{equation}
{\cal N}(4,1){=}...{=}{\cal N}(4,5){=}0, \ {\cal N}(4,6){=}6
,
\label{eq112b}
\end{equation}
and so on.  
See the list of all recursive relations for $d<100$ with initial conditions 
in the Supplementary Material \cite{supplementary}. 
The corresponding large $n$ asymptotics are 
\begin{align}
{\cal N}(1,n) &\cong z_a^n
,
\nonumber
\\[3pt]
{\cal N}(2,n) &\cong \frac{1}{z_a(2z_a^2 + 3z_a + 4)}\, n z_a^{n} = 0.05376 n z_a^n
,
\nonumber
\\[3pt]
{\cal N}(4,n) &\cong  \frac{1}{2!}\Bigl[ \frac{1}{z_a(2z_a^2 + 3z_a + 4)} \Bigr]^2 n^2 z_a^n
= 0.001445 n^2 z_a^n
,
\label{eq117}
\end{align}
and so on. See the list of the resulting large $n$ asymptotics of ${\cal N}(d{<}100,n)$  
in the Supplementary Material \cite{supplementary}.

More generally, for any value of degeneracy $d$ one can write the recursion relation for ${\cal N}(d,n)$ in terms of the multiplicative partition of $d$ into prime degeneracies. 
Let us present the full set of recursion relations for ${\cal N}(d,n)$. The reasoning is as follows. 

(i) The recursion relation for ${\cal N}(1,n)$ is given by Eq.~(\ref{eq90}) with the initial condition Eq.~(\ref{eq105}).  

(ii) For the largest degeneracies, ${\cal N}(d_D(n),n)=1$. 

(iii) All prime degeneracies $\tilde{d}(\ell)$ except $\tilde{d}(5)=4$ are not multiplicatively separable into other prime degeneracies (primality property). The only exception is $\tilde{d}(5)=\tilde{d}^2(4)= 2^2$. See Sec.~\ref{s3pp} above for more detail. 

(iv) Any degeneracy in the spectrum except $d_D$ (and $d=1$) is 
multiplicatively separable into the prime degeneracies $\tilde{d}(\ell)$, namely, 
\begin{align}
d &= \tilde{d}^{\,\mu_4}(4)\tilde{d}^{\,\mu_5}(5)\tilde{d}^{\,\mu_6}(6)\tilde{d}^{\,\nu_7}(7) 
... \nonumber\\
&= 2^{\,\mu_4} 4^{\,\mu_5} 7^{\,\mu_6} 12^{\,\mu_7} 
... \nonumber\\
&= 2^{\,\mu_4+2\mu_5}  7^{\,\mu_6} 12^{\,\mu_7} 
...
, 
\label{a200}
\end{align}
where the powers $\mu_\ell \equiv \mu(\ell)$ are non-negative integers. 
Due to the exception $\tilde{d}(5)=\tilde{d}^2(4)= 2^2$ and the coincidence $\tilde{d}(5)\tilde{d}(8)=4\times21=7\times12=\tilde{d}(6)\tilde{d}(7)$, etc., the multiplicative partition into the prime degeneracies $\tilde{d}(\ell)$, Eq.~(\ref{a200}), 
may not be unique, see below. 
With all possible integers $\mu(4)\geq 0$, 
$\mu(6)\geq 0$, $\mu(7)\geq 0$ ... we generate the full list ${\cal D}$ of all degeneracies except $1$ and $d_D$ in the spectrum for $n\to\infty$. 
We place the generated degeneracies in ascending order, ignoring repetitions for some degeneracies. 
For a finite $n$, only some on the degeneracies from this list are present in the spectrum. In practice, we need degeneracies $d$ only up to some, maybe, large but finite value. So, generating this list, we use a finite set of non-negative integers $\mu_4,\mu_6,\mu_7,...$ respectively restricted from above. 

(v) For any degeneracy $d$ in this list ${\cal D}$ we define the vector ${\cal L}(d) \equiv (\ell_1,\ell_2,...,\ell_\text{max})$, indicating that the prime degeneracies $(\tilde{d}(\ell_1),\tilde{d}(\ell_2),...,\tilde{d}(\ell_\text{max}))$, given in ascending order, are present in at least one of 
the partitions of $d$, Eq.~(\ref{a200}). 
That is, if a prime partition $\tilde{d}(\ell)$ is present in ${\cal L}(d)$, then the ratio $d/\tilde{d}(\ell)$ also belongs to ${\cal D}$.
Then the recursion relation for ${\cal N}(d,n)$ has the form 
\begin{multline}
{\cal N}(d,n) = {\cal N}(d,n{-}2) + {\cal N}(d,n{-}3) + {\cal N}(d,n{-}4) 
\\ + \sum_{\ell \in {\cal L}(d)} {\cal N}(d/\tilde{d}(\ell),n{-}\ell{-}1)
, 
\label{a210}
\end{multline}
where $\ell$ in the sum takes the values $\ell_1,\ell_2,...,\ell_\text{max}$, i.e., the components of the vector ${\cal L}(d)$. 
This formula sums the number of configurations ${\cal N}(d/\tilde{d}(\ell),n{-}\ell{-}1)$ of smaller systems to which a block of $\ell$ zeros (followed by a one) can be added in order to form a configuration of size $n$ and degeneracy $d$.
We should not explicitly include in this sum configurations that achieve degeneracy $d$ by inserting more than one block of zeros. The insertion of additional blocks besides a block of length $l$, say, into smaller configurations are already accounted for in the calculation of the degeneracy of the configuration of size $n{-}\ell{-}1$ into which the block of length $l$ will be inserted.
Note that the ratio $d/\tilde{d}(\ell)$ in each of the terms in the sum is one of the degeneracies, smaller than $d$, in the list ${\cal D}$ with the degeneracy $d=1$ added, so this set of recursions together with the recursion for ${\cal N}(1,n)$, Eq.~(\ref{eq90}), is closed.   
The initial conditions for these recursions are 
\begin{align}
{\cal N}(d,1) &= ... = {\cal N}(d,\ell_\text{max}) = 0, \nonumber\\
{\cal N}(d,\ell_\text{max}{+}1) &= (\ell_\text{max}{+}1)\delta(d,\tilde{d}(\ell_\text{max}))
, 
\label{a220}
\end{align}
where $\delta(i,j)$ is the Kronecker symbol. 
Note than when $d$ is a prime degeneracy $d=\tilde{d}(\ell)=\tilde{d}(\ell_\text{max})$ except $d=4=\tilde{d}(5)$, the sum on the left-hand side of Eq.~(\ref{a210}) has only one term, ${\cal N}(1,n{-}\ell_\text{max}{-}1)$. 
If $d=4$, Eqs.~(\ref{a210}) and (\ref{a220}) properly give the recursion for from Eq.~(\ref{eq112a}) with the initial condition from Eq.~(\ref{eq112b}). Finally, for $d=1$, use Eq.~(\ref{eq90}) with the initial condition Eq.~(\ref{eq105}). [In fact, if $d=1$, the list ${\cal L}(d=1)$ is empty, and Eq.~(\ref{a210}) is reduced to Eq.~(\ref{eq90}). We still need the initial condition, Eq.~(\ref{eq105}), for $d=1$, since Eq.~(\ref{a220}) is not defined in this case.]

(vi) Equations~(\ref{eq90}), (\ref{eq105}) and (\ref{a210}), (\ref{a220}), in particular, provide all degeneracies present in the spectrum for a given $n$. If some $d$ is absent, these recursions produce ${\cal N}(d,n)=0$. 
For instance, let $d=64$. The initial condition is ${\cal N}(64,1){=}...{=}{\cal N}(64,6){=}0$. The recursions produce ${\cal N}(64,7){=}...{=}{\cal N}(64,17){=}0$, ${\cal N}(64,18){=}6$, ${\cal N}(64,19){=}0$, ${\cal N}(64,20){=}20$, 
and so on. 

(vii) Our recursions, Eq.~(\ref{a210}), with their initial conditions, Eq.~(\ref{a220}), lead to the following large $n$ asymptotics of ${\cal N}(d,n)$: 
\begin{multline}
{\cal N}(d,n) \\
\cong \frac{1}{\prod_{\ell\neq5} \mu_\ell !} \Bigl[ \frac{n}{z_a(2z_a^2 + 3z_a + 4)} \Bigr]^{\sum_{\ell\neq5} \mu_\ell}  z_a^{n - \sum_{\ell\geq6}(\ell-4)\mu_\ell}
\label{a230}
\end{multline}
for $d$ with the multiplicative partition into prime degeneracies, Eq.~(\ref{a210}), chosen in such a way that the power of $2$, $\mu_4$, in this partition is maximal. 

Let us discuss this point in more detail. We stressed above that a multiplicative partition of $d\in{\cal D}$, Eq.~(\ref{a210}), may not be unique. For large $n$, the contribution of each of the possible multiplicative partitions of $d$ to ${\cal N}(d,n)$ is about $n^{\sum_{\ell} \mu_\ell}  z_a^{n - \sum_{\ell}(\ell-4)\mu_\ell}$. 
The leading asymptotics, Eq.~(\ref{a230}), is with the maximal power of $n$, i.e., it originates from the partition with the maximal sum $\sum_\ell \mu_\ell$. 
Let us check whether the multiplicative partition of $d$ with the maximal sum $\sum_\ell \mu_\ell$ is unique and that this maximum corresponds to the maximal $\mu_4$. In other words, if there exist a number of different multiplicative partitions of $d$, that only one of them has the maximal $\mu_4$, and that this partition has the maximal sum $\sum_\ell \mu_\ell$.

To explain why the property of primality is very likely to hold for all large enough $\tilde{d}(\ell)$, let us consider the number of conventional prime numbers smaller than some number $d$, which grows as $\sim d/\ln{d}$. 
Since $\tilde{d}(\ell) \sim z_d^\ell$, the number of conventional primes smaller than $\tilde{d}(\ell)$ grows exponentially with $\ell$ as $\sim z_d^\ell/\ell$. 
Additionally, the average number of conventional primes in the factorization of numbers of the magnitude of $d$ grows as $\sim \ln \ln d$~\cite{erdos1940gaussian}. 
A necessary condition for some $\tilde{d}(\ell)$ to not be prime is that all of its prime factors are also factors of at least another $\tilde{d} < \tilde{d}(\ell)$.
The combination of the exponential increase of conventional primes smaller than $\tilde{d}(\ell)$, and the increase of the number of factors of $\tilde{d}(\ell)$ as $\ell$ increases, makes the probability of the degeneracy $\tilde{d}(\ell)$ not being prime approach $0$ very rapidly.
In the set of values $\tilde{d}(\ell \leq 200)$, the only ones that do not contain at least one conventional prime factor absent from all factorizations of smaller $\tilde{d}$ are $\tilde{d}(5)=4$, $\tilde{d}(8)=21$, $\tilde{d}(12)=200$, $\tilde{d}(13)=351$, and $\tilde{d}(24)=170\, 625$, and from these only $\tilde{d}(5)$ is actually not a prime degeneracy.

We inspected all products of prime degeneracies 
$\tilde{d}(\ell\leq200)\approx 1.7\, 10^{48}$  
e focusing on the products producing non-unique multiplicative partitions. Apart from $2^2=4$ discussed above, we find only three such combinations with $\ell\leq200$ (into this number, we do not include the products of these combinations and arbitrary prime degeneracies). Namely, $2^2 \times21=7\times 12$, $2^9\times 200\times 351^2 = 12^6\times 65^2$, and  $2^{2}\times  12^{2}\times 170625 =7 \times 200^2 \times 351$. For each of these combinations we confirm that the left side  of the equality corresponds to the maximal $\sum_\ell \mu_\ell$ and that this partitions is unique. Clearly, the same is true for the products of these combinations and arbitrary prime degeneracies. Thus partitions contributing to the leading asymptotics of ${\cal N}(d,n)$ are unique, and the gauge is fixed by demanding $\mu_4=\text{max}$, which leads to Eq.~(\ref{a230}).  
For grasping the form of Eq.~(\ref{a230}), see also Eqs.~(\ref{eq117}), especially the asymptotics of ${\cal N}(84,n)$.

Thus we have the chain of conveniently coupled linear recursion relations that one can easily process starting from $d=1$. These recursions generate exact ${\cal N}(d,n)$ for finite $d$ and $n$, and in this sense provide the exact solution of the problem. 
Moreover, the leading large $n$ asymptotics of ${\cal N}(d,n)$ are found explicitly. 
Note that the time of computing all ${\cal N}(d<d_0,n)$, where $d_0$ is fixed, by using our recursions, is proportional to $n$, i.e. the computation for each next size takes the same time.

\section{Moments of ${\cal N}(d,n)$}
\label{a3}

Here we list the large $n$ asymptotics of the moments of ${\cal N}(d,n)$. 
The leading asymptotics of the moments are of the following form:   
\begin{equation}
{\cal M}_k(n) \equiv \sum_{i=1}^D d_i^k {\cal N}(d_i,n) \cong c_k^n 
,  
\label{a20010}
\end{equation}
where 
\begin{eqnarray}
c_0 & = & z_g 
, 
\nonumber 
\\ 
c_1 & = & 2 
, 
\nonumber 
\\ 
c_2 & = & 3.13899009933542 
, 
\nonumber 
\\ 
c_3 & = & 5.41762864130976 
, 
\nonumber 
\\ 
c_4 & = & 9.48696631140060 
, 
\nonumber 
\\ 
c_5 & = & 16.6438119672308 
, 
\nonumber 
\\ 
c_6 & = & 29.2067717071942 
, 
\nonumber 
\\ 
c_7 & = & 51.2540583046806 
, 
\nonumber 
\\ 
c_8 & = & 89.9445429351823 
, 
\nonumber 
\\ 
& ... &
\label{a20020}
\end{eqnarray}
In their turn, the numbers $c_k$ have the following large $k$ asymptotics: 
\begin{equation}
c_k \cong z_d^k [1 + (z_d^4 - 2)^{-k} + 
 ...]
,   
\label{a20030}
\end{equation}
which is close to the numerical values listed above, Eq.~(\ref{a20020}), already for $k=2$. The leading term $z_d^k$ results from the largest degeneracy in the spectrum, ${\cal N}(d_D{\cong} z_d^n,n)=1$, the next term originates from the second largest degeneracy, etc.

\section{Asymptotic mean-field degeneracy distribution}\label{a5}

Let us obtain an explicit formula for the asymptotics of the mean-field approximation to the degeneracy distribution $\overline{\cal N}$ in terms 
of the degeneracy $d$. We simply replace $m$ in Eq.~(\ref{eq50}) by the inverse of the function $\overline{d}(m)$ in Eq.~(\ref{eq40}). 
The inverse function is $m(d)=(\ln d - n \ln z_d)/(\ln C_1- \ln z_d)$ (where we have dropped the bar because $d$ is now the independent variable), and substituting into Eq.~(\ref{eq50}) gives
\begin{eqnarray}
\overline{\cal N} & = & \frac{n}{n-m} {{n-m}\choose{m}} \approx \frac{n^m}{m!} 
\nonumber
\\[3pt]
& = & 
\frac{\frac{n \ln n}{\ln (z_d/C_1)}}{\Gamma\left[ 1 - \frac{\ln (d/z_d^n)}{\ln (z_d/C_1)} \right]}   d^{-\frac{\ln n}{\ln (z_d/C_1)}} 
\label{eq60}
\end{eqnarray}
for $m\ll n$. 
Note that the logarithmic derivative of $\overline{\cal N}$ over $d$ at the point $d_D$ (the highest degeneracy), 
$-\ln n/\ln(z_d^4-2)$, properly fits the two rightmost points of the spectrum, namely $d_D(n) \cong z_d^n$, ${\cal N}(d_D) = 1$ and $d_{D-1}(n) \cong z_d^n/(z_d^4 - 2)$, 
${\cal N}(d_{D-1}) = n$.



%











\setcounter{section}{0}
\setcounter{equation}{0}
\setcounter{figure}{0}
\setcounter{table}{0}
\setcounter{page}{1}
\makeatletter

\renewcommand{\thetable}{S\arabic{table}}
\renewcommand{\theequation}{S\arabic{equation}}
\renewcommand{\thefigure}{S\arabic{figure}}
\renewcommand{\bibnumfmt}[1]{[S#1]}
\renewcommand{\citenumfont}[1]{S#1}
\renewcommand\thesection{\Roman{section}}. 
\renewcommand\appendixname{S}

\pagebreak
\newpage
\clearpage

\widetext

\begin{center}
\textbf{\large Supplemental Materials: Complex distributions emerging in filtering and compression}
\end{center}

\section{RECURSIONS AND ASYMPTOTICS FOR ${\cal N}(d<100,n)$}
\label{supp1}

Here we show the recursion relations for ${\cal N}(d,n)$ with degeneracy $d$ up to $100$, initial conditions for them, and their large $n$ asymptotics. These formulas are particular cases of Eqs.~(A12), (A13), and (A14), respectively. 
\begin{eqnarray}
&&
{\cal N}(1,n) = {\cal N}(1,n{-}2) + {\cal N}(1,n{-}3) + {\cal N}(1,n{-}4) 
,
\nonumber
\\[3pt]
&&
{\cal N}(2,n) = {\cal N}(2,n{-}2) + {\cal N}(2,n{-}3) + {\cal N}(2,n{-}4)  +  {\cal N}(1,n{-}5)
,
\nonumber
\\[3pt]
&&
{\cal N}(4,n) = {\cal N}(4,n{-}2) + {\cal N}(4,n{-}3) + {\cal N}(4,n{-}4)  + {\cal N}(2,n{-}5) + {\cal N}(1,n{-}6)
,
\nonumber
\\[3pt]
&&
{\cal N}(7,n) = {\cal N}(7,n{-}2) + {\cal N}(7,n{-}3) + {\cal N}(7,n{-}4) + {\cal N}(1,n{-}7)
,
\nonumber
\\[3pt]
&&
{\cal N}(8,n) = {\cal N}(8,n{-}2) + {\cal N}(8,n{-}3) + {\cal N}(8,n{-}4) + {\cal N}(4,n{-}5) + {\cal N}(2,n{-}6)
,
\nonumber
\\[3pt]
&&
{\cal N}(12,n) = {\cal N}(12,n{-}2) + {\cal N}(12,n{-}3) + {\cal N}(12,n{-}4) + {\cal N}(1,n{-}8)
,
\nonumber
\\[3pt]
&&
{\cal N}(14,n) = {\cal N}(14,n{-}2) + {\cal N}(14,n{-}3) + {\cal N}(14,n{-}4) + {\cal N}(7,n{-}5) + {\cal N}(2,n{-}7)
,
\nonumber
\\[3pt]
&&
{\cal N}(16,n) = {\cal N}(16,n{-}2) + {\cal N}(16,n{-}3) + {\cal N}(16,n{-}4) + {\cal N}(8,n{-}5) + {\cal N}(4,n{-}6)
,
\nonumber
\\[3pt]
&&
{\cal N}(21,n) = {\cal N}(21,n{-}2) + {\cal N}(21,n{-}3) + {\cal N}(21,n{-}4) + {\cal N}(1,n{-}9)
,
\nonumber
\\[3pt]
&&
{\cal N}(24,n) = {\cal N}(24,n{-}2) + {\cal N}(24,n{-}3) + {\cal N}(24,n{-}4) + {\cal N}(12,n{-}5) + {\cal N}(2,n{-}8)
,
\nonumber
\\[3pt]
&&
{\cal N}(28,n) = {\cal N}(28,n{-}2) + {\cal N}(28,n{-}3) + {\cal N}(28,n{-}4) + {\cal N}(14,n{-}5) + {\cal N}(7,n{-}6)+ {\cal N}(4,n{-}7)
,
\nonumber
\\[3pt]
&&
{\cal N}(32,n) = {\cal N}(32,n{-}2) + {\cal N}(32,n{-}3) + {\cal N}(32,n{-}4) + {\cal N}(16,n{-}5) + {\cal N}(8,n{-}6)
,
\nonumber
\\[3pt]
&&
{\cal N}(37,n) = {\cal N}(37,n{-}2) + {\cal N}(37,n{-}3) + {\cal N}(37,n{-}4) + {\cal N}(1,n{-}10)
,
\nonumber
\\[3pt]
&&
{\cal N}(42,n) = {\cal N}(42,n{-}2) + {\cal N}(42,n{-}3) + {\cal N}(42,n{-}4) + {\cal N}(21,n{-}5) + {\cal N}(2,n{-}9)
,
\nonumber
\\[3pt]
&&
{\cal N}(48,n) = {\cal N}(48,n{-}2) + {\cal N}(48,n{-}3) + {\cal N}(48,n{-}4) + {\cal N}(24,n{-}5) + {\cal N}(12,n{-}6)+ {\cal N}(4,n{-}8)
,
\nonumber
\\[3pt]
&&
{\cal N}(49,n) = {\cal N}(49,n{-}2) + {\cal N}(49,n{-}3) + {\cal N}(49,n{-}4) + {\cal N}(7,n{-}7)
,
\nonumber
\\[3pt]
&&
{\cal N}(56,n) = {\cal N}(56,n{-}2) + {\cal N}(56,n{-}3) + {\cal N}(56,n{-}4) + {\cal N}(28,n{-}5) + {\cal N}(14,n{-}6)+ {\cal N}(8,n{-}7)
,
\nonumber
\\[3pt]
&&
{\cal N}(64,n) = {\cal N}(64,n{-}2) + {\cal N}(64,n{-}3) + {\cal N}(64,n{-}4) + {\cal N}(32,n{-}5) + {\cal N}(16,n{-}6)
,
\nonumber
\\[3pt]
&&
{\cal N}(65,n) = {\cal N}(65,n{-}2) + {\cal N}(65,n{-}3) + {\cal N}(65,n{-}4) + {\cal N}(1,n{-}11)
,
\nonumber
\\[3pt]
&&
{\cal N}(74,n) = {\cal N}(74,n{-}2) + {\cal N}(74,n{-}3) + {\cal N}(74,n{-}4) + {\cal N}(37,n{-}5) + {\cal N}(2,n{-}10)
,
\nonumber
\\[3pt]
&&
{\cal N}(84,n) = {\cal N}(84,n{-}2) {+} {\cal N}(84,n{-}3) {+} {\cal N}(84,n{-}4) {+} {\cal N}(42,n{-}5) {+} {\cal N}(21,n{-}6) {+} {\cal N}(12,n{-}7) {+} {\cal N}(7,n{-}8) {+} {\cal N}(4,n{-}9)
,
\nonumber
\\[3pt]
&&
{\cal N}(96,n) = {\cal N}(96,n{-}2) + {\cal N}(96,n{-}3) + {\cal N}(96,n{-}4) + {\cal N}(48,n{-}5) + {\cal N}(24,n{-}6)+ {\cal N}(8,n{-}8)
,
\nonumber
\\[3pt]
&&
{\cal N}(98,n) = {\cal N}(98,n{-}2) + {\cal N}(98,n{-}3) + {\cal N}(98,n{-}4) + {\cal N}(49,n{-}5) + {\cal N}(14,n{-}7)
\label{su10}
\end{eqnarray}
with the initial conditions 
\begin{eqnarray}
&&
{\cal N}(1,1){=}0, \  {\cal N}(1,2){=}2, \ {\cal N}(1,3){=}3, \ {\cal N}(1,4){=}6
,
\nonumber
\\[3pt]
&&
{\cal N}(2,1){=}...{=}{\cal N}(2,4){=}0, \  {\cal N}(2,5){=}5
,
\nonumber
\\[3pt]
&&
{\cal N}(4,1){=}...{=}{\cal N}(4,5){=}0, \ {\cal N}(4,6){=}6
,
\nonumber
\\[3pt]
&&
{\cal N}(7,1){=}...{=}{\cal N}(7,6){=}0, \ {\cal N}(7,7){=}7
,
\nonumber
\\[3pt]
&&
{\cal N}(8,1){=}...{=}{\cal N}(8,6){=}0
,
\nonumber
\\[3pt]
&&
{\cal N}(12,1){=}...{=}{\cal N}(12,7){=}0, \ {\cal N}(12,8){=}8
,
\nonumber
\\[3pt]
&&
{\cal N}(14,1){=}...{=}{\cal N}(14,7){=}0
,
\nonumber
\\[3pt]
&&
{\cal N}(16,1){=}...{=}{\cal N}(16,6){=}0
,
\nonumber
\\[3pt]
&&
{\cal N}(21,1){=}...{=}{\cal N}(21,8){=}0, \ {\cal N}(21,9){=}9
,
\nonumber
\\[3pt]
&&
{\cal N}(24,1){=}...{=}{\cal N}(24,8){=}0
,
\nonumber
\\[3pt]
&&
{\cal N}(28,1){=}...{=}{\cal N}(28,7){=}0
,
\nonumber
\\[3pt]
&&
{\cal N}(32,1){=}...{=}{\cal N}(32,6){=}0
,
\nonumber
\\[3pt]
&&
{\cal N}(37,1){=}...{=}{\cal N}(37,9){=}0, \ {\cal N}(37,10){=}10
,
\nonumber
\\[3pt]
&&
{\cal N}(42,1){=}...{=}{\cal N}(42,9){=}0
,
\nonumber
\\[3pt]
&&
{\cal N}(48,1){=}...{=}{\cal N}(48,8){=}0
,
\nonumber
\\[3pt]
&&
{\cal N}(49,1){=}...{=}{\cal N}(49,7){=}0
,
\nonumber
\\[3pt]
&&
{\cal N}(56,1){=}...{=}{\cal N}(56,7){=}0
,
\nonumber
\\[3pt]
&&
{\cal N}(64,1){=}...{=}{\cal N}(64,6){=}0
,
\nonumber
\\[3pt]
&&
{\cal N}(65,1){=}...{=}{\cal N}(65,10){=}0, \ {\cal N}(65,11){=}11
,
\nonumber
\\[3pt]
&&
{\cal N}(74,1){=}...{=}{\cal N}(74,10){=}0
,
\nonumber
\\[3pt]
&&
{\cal N}(84,1){=}...{=}{\cal N}(84,9){=}0
,
\nonumber
\\[3pt]
&&
{\cal N}(96,1){=}...{=}{\cal N}(96,8){=}0
,
\nonumber
\\[3pt]
&&
{\cal N}(98,1){=}...{=}{\cal N}(98,7){=}0
.
\label{su20}
\end{eqnarray}
The corresponding large $n$ asymptotics are 
\begin{eqnarray}
&&
{\cal N}(1,n) \cong z_a^n
,
\nonumber
\\[3pt]
&&
{\cal N}(2,n) \cong \frac{1}{z_a(2z_a^2 + 3z_a + 4)}\, n z_a^{n} = 0.05376 n z_a^n
,
\nonumber
\\[3pt]
&&
{\cal N}(4,n) \cong  \frac{1}{2!}\Bigl[ \frac{1}{z_a(2z_a^2 + 3z_a + 4)} \Bigr]^2 n^2 z_a^n
= 0.001445 n^2 z_a^n
,
\nonumber
\\[3pt]
&&
{\cal N}(7,n) \cong   \frac{1}{z_a(2z_a^2 + 3z_a + 4)}\, n z_a^{n-2}
= 0.02503 n z_a^n
,
\nonumber
\\[3pt]
&&
{\cal N}(8,n) \cong  \frac{1}{3!}\Bigl[ \frac{1}{z_a(2z_a^2 + 3z_a + 4)} \Bigr]^3 n^3 z_a^n
=  0.00002589 n^3 z_a^n
,
\nonumber
\\[3pt]
&&
{\cal N}(12,n) \cong  \frac{1}{z_a(2z_a^2 + 3z_a + 4)}\, n z_a^{n-3}
= 0.01708 n z_a^n
,
\nonumber
\\[3pt]
&&
{\cal N}(14,n) \cong \Bigl[ \frac{1}{z_a(2z_a^2 + 3z_a + 4)} \Bigr]^2 n^2 z_a^{n-2}
= 0.001345 n^2 z_a^n
,
\nonumber
\\[3pt]
&&
{\cal N}(16,n) \cong \frac{1}{4!}\Bigl[ \frac{1}{z_a(2z_a^2 + 3z_a + 4)} \Bigr]^4 n^4 z_a^n 
= 3.480{\times} 10^{-7} n^4 z_a^n
,
\nonumber
\\[3pt]
&&
{\cal N}(21,n) \cong  \frac{1}{z_a(2z_a^2 + 3z_a + 4)}\, n z_a^{n-4}
,
\nonumber
\\[3pt]
&&
{\cal N}(24,n) \cong \Bigl[ \frac{1}{z_a(2z_a^2 + 3z_a + 4)} \Bigr]^2 n^2 z_a^{n-3}
,
\nonumber
\\[3pt]
&&
{\cal N}(28,n) \cong  \frac{1}{2!} \Bigl[ \frac{1}{z_a(2z_a^2 + 3z_a + 4)} \Bigr]^3 n^3 z_a^{n-2}
,
\nonumber
\\[3pt]
&&
{\cal N}(32,n) \cong \frac{1}{5!}\Bigl[ \frac{1}{z_a(2z_a^2 + 3z_a + 4)} \Bigr]^5 n^5 z_a^n 
,
\nonumber
\\[3pt]
&&
{\cal N}(37,n) \cong  \frac{1}{z_a(2z_a^2 + 3z_a + 4)}\, n z_a^{n-5}
,
\nonumber
\\[3pt]
&&
{\cal N}(42,n) \cong \Bigl[ \frac{1}{z_a(2z_a^2 + 3z_a + 4)} \Bigr]^2 n^2 z_a^{n-4}
,
\nonumber
\\[3pt]
&&
{\cal N}(48,n) \cong \frac{1}{2!} \Bigl[ \frac{1}{z_a(2z_a^2 + 3z_a + 4)} \Bigr]^3 n^3 z_a^{n-3}
,
\nonumber
\\[3pt]
&&
{\cal N}(49,n) \cong  \frac{1}{2!} \Bigl[ \frac{1}{z_a(2z_a^2 + 3z_a + 4)} \Bigr]^2 n^2 z_a^{n-4}
,
\nonumber
\\[3pt]
&&
{\cal N}(56,n) \cong  \frac{1}{3!} \Bigl[ \frac{1}{z_a(2z_a^2 + 3z_a + 4)} \Bigr]^4 n^4 z_a^{n-2}
,
\nonumber
\\[3pt]
&&
{\cal N}(64,n) \cong \frac{1}{6!}\Bigl[ \frac{1}{z_a(2z_a^2 + 3z_a + 4)} \Bigr]^6 n^6 z_a^n 
,
\nonumber
\\[3pt]
&&
{\cal N}(65,n) \cong  \frac{1}{z_a(2z_a^2 + 3z_a + 4)}\, n z_a^{n-6}
,
\nonumber
\\[3pt]
&&
{\cal N}(74,n) \cong \Bigl[ \frac{1}{z_a(2z_a^2 + 3z_a + 4)} \Bigr]^2 n^2 z_a^{n-5}
,
\nonumber
\\[3pt]
&&
{\cal N}(84,n) \cong  \frac{1}{2!} \Bigl[ \frac{1}{z_a(2z_a^2 + 3z_a + 4)} \Bigr]^3 n^3 z_a^{n-4}
,
\nonumber
\\[3pt]
&&
{\cal N}(96,n) \cong  \frac{1}{3!} \Bigl[ \frac{1}{z_a(2z_a^2 + 3z_a + 4)} \Bigr]^4 n^4 z_a^{n-3}
,
\nonumber
\\[3pt]
&&
{\cal N}(98,n) \cong  \frac{1}{2!} \Bigl[ \frac{1}{z_a(2z_a^2 + 3z_a + 4)} \Bigr]^3 n^3 z_a^{n-4}
,
\nonumber
\\[3pt]
&& \hspace{41pt}
... \, 
.
\label{su30}
\end{eqnarray}


\section{TABLES OF EXACT RESULTS FOR COMPLETE INPUT DATA SETS}
\label{supp2}

Table~\ref{t1} contains the complete data on the spectrum of degeneracy of outputs for complete input data sets of length $n$ up to $n=21$, namely  
all pairs $\{ d_i,{\cal N}_i \}$, $i=1,...,D$, where $d_i$ is the $i$-th value of degeneracy and ${\cal N}_i$ is the corresponding number of outputs having this degeneracy. 
This data for $n=21$ is presented in graphical form in Fig.~3~(a,b). 
Table~\ref{t2} presents the total number of different values of degeneracy $D$ for each input length $n\leq 120$. 



\begin{table*}[hb]
\caption{Spectrum of degeneracy produced by complete input data sets with input strings of length $n$, i.e., the pairs $\{d_i,{\cal N}_i\}$, $i = 1,...,D$, where $d_i$ is the $i$-th value of degeneracy and ${\cal N}_i$ is the corresponding number of outputs having this degeneracy, $D$ is the total number of different values of degeneracy.} 
{\scriptsize  
\begin{tabular} 
{c
l} 
\hline 
&  
\\[-6pt] 
$n$
&  
\hspace{230pt} $\genfrac{}{}{0pt}{0}{d_j}{{\cal N}_j}$
\\[7pt]  
\hline 
&   
\\[-6pt] 
3 
& 
$\genfrac{}{}{0pt}{0}{1}{3}\ \genfrac{}{}{0pt}{0}{5}{1}$   
\\[5pt] 
\hline 
&   
\\[-6pt]
4
&  
$\genfrac{}{}{0pt}{0}{1}{6}\ \genfrac{}{}{0pt}{0}{10}{1}$      
\\[5pt] 
\hline 
&   
\\[-6pt]
5 
& 
$\genfrac{}{}{0pt}{0}{1}{5}\ \genfrac{}{}{0pt}{0}{2}{5}\ \genfrac{}{}{0pt}{0}{17}{1}$     
\\[5pt] 
\hline 
&   
\\[-6pt]
6
&  
$\genfrac{}{}{0pt}{0}{1}{11}\ \genfrac{}{}{0pt}{0}{4}{6}\ \genfrac{}{}{0pt}{0}{29}{1}$   
\\[5pt] 
\hline
&   
\\[-6pt]
7 
& 
$\genfrac{}{}{0pt}{0}{1}{14}\ \genfrac{}{}{0pt}{0}{2}{7}\ \genfrac{}{}{0pt}{0}{7}{7}\ \genfrac{}{}{0pt}{0}{51}{1}$   
\\[5pt] 
\hline 
&   
\\[-6pt]
8
&  
$\genfrac{}{}{0pt}{0}{1}{22}\ \genfrac{}{}{0pt}{0}{2}{8}\ \genfrac{}{}{0pt}{0}{4}{8}\ \genfrac{}{}{0pt}{0}{12}{8}\ \genfrac{}{}{0pt}{0}{90}{1}$     
\\[5pt] 
\hline
&   
\\[-6pt]
9 
& 
$\genfrac{}{}{0pt}{0}{1}{30}\ \genfrac{}{}{0pt}{0}{2}{18}\ \genfrac{}{}{0pt}{0}{4}{9}\ \genfrac{}{}{0pt}{0}{7}{9}\ \genfrac{}{}{0pt}{0}{21}{9}\ \genfrac{}{}{0pt}{0}{158}{1}$   
\\[5pt] 
\hline 
&   
\\[-6pt]
10
&  
$\genfrac{}{}{0pt}{0}{1}{47}\ \genfrac{}{}{0pt}{0}{2}{20}\ \genfrac{}{}{0pt}{0}{4}{25}\ \genfrac{}{}{0pt}{0}{7}{10}\ \genfrac{}{}{0pt}{0}{12}{10}\ \genfrac{}{}{0pt}{0}{37}{10}\ \genfrac{}{}{0pt}{0}{277}{1}$     
\\[5pt] 
\hline
&   
\\[-6pt]
11 
& 
$\genfrac{}{}{0pt}{0}{1}{66}\ \genfrac{}{}{0pt}{0}{2}{44}\ \genfrac{}{}{0pt}{0}{4}{22}\ \genfrac{}{}{0pt}{0}{7}{22}\ \genfrac{}{}{0pt}{0}{8}{11}\ \genfrac{}{}{0pt}{0}{12}{11}\ \genfrac{}{}{0pt}{0}{21}{11}\ \genfrac{}{}{0pt}{0}{65}{11}\ \genfrac{}{}{0pt}{0}{486}{1}$       
\\[5pt] 
\hline 
&   
\\[-6pt]
12
&  
$\genfrac{}{}{0pt}{0}{1}{99}\ \genfrac{}{}{0pt}{0}{2}{60}\ \genfrac{}{}{0pt}{0}{4}{60}\ \genfrac{}{}{0pt}{0}{7}{24}\ \genfrac{}{}{0pt}{0}{12}{24}\ \genfrac{}{}{0pt}{0}{14}{12}\ \genfrac{}{}{0pt}{0}{16}{6}\ \genfrac{}{}{0pt}{0}{21}{12}\ \genfrac{}{}{0pt}{0}{37}{12}\ \genfrac{}{}{0pt}{0}{114}{12}\ \genfrac{}{}{0pt}{0}{853}{1}$     
\\[5pt] 
\hline
&   
\\[-6pt]
13 
& 
$\genfrac{}{}{0pt}{0}{1}{143}\ \genfrac{}{}{0pt}{0}{2}{104}\ \genfrac{}{}{0pt}{0}{4}{78}\ \genfrac{}{}{0pt}{0}{7}{52}\ \genfrac{}{}{0pt}{0}{8}{26}\ \genfrac{}{}{0pt}{0}{12}{26}\ \genfrac{}{}{0pt}{0}{21}{26}\ \genfrac{}{}{0pt}{0}{24}{13}\ \genfrac{}{}{0pt}{0}{28}{13}\ \genfrac{}{}{0pt}{0}{37}{13}\ \genfrac{}{}{0pt}{0}{65}{13}\ \genfrac{}{}{0pt}{0}{200}{13}\ \genfrac{}{}{0pt}{0}{1497}{1}$    
\\[5pt] 
\hline 
&   
\\[-6pt]
14
&  
$\genfrac{}{}{0pt}{0}{1}{212}\ \genfrac{}{}{0pt}{0}{2}{154}\ \genfrac{}{}{0pt}{0}{4}{147}\ \genfrac{}{}{0pt}{0}{7}{70}\ \genfrac{}{}{0pt}{0}{8}{28}\ \genfrac{}{}{0pt}{0}{12}{56}\ \genfrac{}{}{0pt}{0}{14}{28}\ \genfrac{}{}{0pt}{0}{16}{14}\ \genfrac{}{}{0pt}{0}{21}{28}\ \genfrac{}{}{0pt}{0}{37}{28}\ \genfrac{}{}{0pt}{0}{42}{14}\ \genfrac{}{}{0pt}{0}{48}{14}\ \genfrac{}{}{0pt}{0}{49}{7}\ \genfrac{}{}{0pt}{0}{65}{14}\ \genfrac{}{}{0pt}{0}{114}{14}\ \genfrac{}{}{0pt}{0}{351}{14}\ \genfrac{}{}{0pt}{0}{2627}{1}$      
\\[5pt] 
\hline
&   
\\[-6pt]
15 
& 
$\genfrac{}{}{0pt}{0}{1}{308}\ \genfrac{}{}{0pt}{0}{2}{255}\ \genfrac{}{}{0pt}{0}{4}{210}\ \genfrac{}{}{0pt}{0}{7}{120}\ \genfrac{}{}{0pt}{0}{8}{80}\ \genfrac{}{}{0pt}{0}{12}{75}\ \genfrac{}{}{0pt}{0}{14}{30}\ \genfrac{}{}{0pt}{0}{16}{15}\ \genfrac{}{}{0pt}{0}{21}{60}\ \genfrac{}{}{0pt}{0}{24}{30}\ \genfrac{}{}{0pt}{0}{28}{30}\ \genfrac{}{}{0pt}{0}{37}{30}\ \genfrac{}{}{0pt}{0}{65}{30}\ \genfrac{}{}{0pt}{0}{74}{15}\ \genfrac{}{}{0pt}{0}{84}{30}\ \genfrac{}{}{0pt}{0}{114}{15}\ \genfrac{}{}{0pt}{0}{200}{15}\ \genfrac{}{}{0pt}{0}{616}{15}\ \genfrac{}{}{0pt}{0}{4610}{1}$    
\\[5pt] 
\hline 
&   
\\[-6pt]
16
&  
$\genfrac{}{}{0pt}{0}{1}{454}\ \genfrac{}{}{0pt}{0}{2}{384}\ \genfrac{}{}{0pt}{0}{4}{376}\ \genfrac{}{}{0pt}{0}{7}{176}\ \genfrac{}{}{0pt}{0}{8}{96}\ \genfrac{}{}{0pt}{0}{12}{128}\ \genfrac{}{}{0pt}{0}{14}{80}\ \genfrac{}{}{0pt}{0}{16}{56}\ \genfrac{}{}{0pt}{0}{21}{80}\ \genfrac{}{}{0pt}{0}{24}{32}\ \genfrac{}{}{0pt}{0}{28}{32}\ \genfrac{}{}{0pt}{0}{37}{64}\ \genfrac{}{}{0pt}{0}{42}{32}\ \genfrac{}{}{0pt}{0}{48}{32}\ \genfrac{}{}{0pt}{0}{49}{16}\ \genfrac{}{}{0pt}{0}{65}{32}\ \genfrac{}{}{0pt}{0}{114}{32}\ \genfrac{}{}{0pt}{0}{130}{16}\ \genfrac{}{}{0pt}{0}{144}{8}\ \genfrac{}{}{0pt}{0}{147}{16}\ \genfrac{}{}{0pt}{0}{148}{16}\ \genfrac{}{}{0pt}{0}{200}{16}\ \genfrac{}{}{0pt}{0}{351}{16}\ \genfrac{}{}{0pt}{0}{1081}{16}\ \genfrac{}{}{0pt}{0}{8090}{1}$      
\\[5pt] 
\hline
&   
\\[-6pt]
17 
& 
$\genfrac{}{}{0pt}{0}{1}{663}\ \genfrac{}{}{0pt}{0}{2}{612}\ \genfrac{}{}{0pt}{0}{4}{561}\ \genfrac{}{}{0pt}{0}{7}{289}\ \genfrac{}{}{0pt}{0}{8}{238}\ \genfrac{}{}{0pt}{0}{12}{187}\ \genfrac{}{}{0pt}{0}{14}{102}\ \genfrac{}{}{0pt}{0}{16}{51}\ \genfrac{}{}{0pt}{0}{21}{136}\ \genfrac{}{}{0pt}{0}{24}{85}\ \genfrac{}{}{0pt}{0}{28}{102}\ \genfrac{}{}{0pt}{0}{32}{17}\ \genfrac{}{}{0pt}{0}{37}{85}\ \genfrac{}{}{0pt}{0}{42}{34}\ \genfrac{}{}{0pt}{0}{48}{34}\ \genfrac{}{}{0pt}{0}{49}{17}\ \genfrac{}{}{0pt}{0}{65}{68}\ \genfrac{}{}{0pt}{0}{74}{34}\ \genfrac{}{}{0pt}{0}{84}{68}\ \genfrac{}{}{0pt}{0}{114}{34}\ \genfrac{}{}{0pt}{0}{200}{34}\ \genfrac{}{}{0pt}{0}{228}{17}\ \genfrac{}{}{0pt}{0}{252}{17}\ \genfrac{}{}{0pt}{0}{259}{17}\ \genfrac{}{}{0pt}{0}{260}{17}\ \genfrac{}{}{0pt}{0}{351}{17}\ \genfrac{}{}{0pt}{0}{616}{17}\ \genfrac{}{}{0pt}{0}{1897}{17}\ \genfrac{}{}{0pt}{0}{14197}{1}$    
\\[5pt] 
\hline 
&   
\\[-6pt]
18
&  
$\genfrac{}{}{0pt}{0}{1}{974}\ \genfrac{}{}{0pt}{0}{2}{936}\ \genfrac{}{}{0pt}{0}{4}{936}\ \genfrac{}{}{0pt}{0}{7}{432}\ \genfrac{}{}{0pt}{0}{8}{342}\ \genfrac{}{}{0pt}{0}{12}{306}\ \genfrac{}{}{0pt}{0}{14}{234}\ \genfrac{}{}{0pt}{0}{16}{171}\ \genfrac{}{}{0pt}{0}{21}{198}\ \genfrac{}{}{0pt}{0}{24}{108}\ \genfrac{}{}{0pt}{0}{28}{108}\ \genfrac{}{}{0pt}{0}{37}{144}\ \genfrac{}{}{0pt}{0}{42}{90}\ \genfrac{}{}{0pt}{0}{48}{108}\ \genfrac{}{}{0pt}{0}{49}{45}\ \genfrac{}{}{0pt}{0}{56}{36}\ \genfrac{}{}{0pt}{0}{64}{6}\ \genfrac{}{}{0pt}{0}{65}{90}\ \genfrac{}{}{0pt}{0}{74}{36}\ \genfrac{}{}{0pt}{0}{84}{72}\ \genfrac{}{}{0pt}{0}{114}{72}\ \genfrac{}{}{0pt}{0}{130}{36}\ \genfrac{}{}{0pt}{0}{144}{18}\ \genfrac{}{}{0pt}{0}{147}{36}\ \genfrac{}{}{0pt}{0}{148}{36}\ \genfrac{}{}{0pt}{0}{200}{36}\ \genfrac{}{}{0pt}{0}{351}{36}\ \genfrac{}{}{0pt}{0}{400}{18}\ \genfrac{}{}{0pt}{0}{441}{9}$   
\\[9pt]   
&  
$\genfrac{}{}{0pt}{0}{444}{18}\ \genfrac{}{}{0pt}{0}{455}{18}\ \genfrac{}{}{0pt}{0}{456}{18}\ \genfrac{}{}{0pt}{0}{616}{18}\ \genfrac{}{}{0pt}{0}{1081}{18}\ \genfrac{}{}{0pt}{0}{3329}{18}\ \genfrac{}{}{0pt}{0}{24914}{1}$
\\[5pt] 
\hline
&   
\\[-6pt]
19 
& 
$\genfrac{}{}{0pt}{0}{1}{1425}\ \genfrac{}{}{0pt}{0}{2}{1463}\ \genfrac{}{}{0pt}{0}{4}{1444}\ \genfrac{}{}{0pt}{0}{7}{684}\ \genfrac{}{}{0pt}{0}{8}{665}\ \genfrac{}{}{0pt}{0}{12}{456}\ \genfrac{}{}{0pt}{0}{14}{342}\ \genfrac{}{}{0pt}{0}{16}{228}\ \genfrac{}{}{0pt}{0}{21}{323}\ \genfrac{}{}{0pt}{0}{24}{247}\ \genfrac{}{}{0pt}{0}{28}{304}\ \genfrac{}{}{0pt}{0}{32}{57}\ \genfrac{}{}{0pt}{0}{37}{209}\ \genfrac{}{}{0pt}{0}{42}{114}\ \genfrac{}{}{0pt}{0}{48}{114}\ \genfrac{}{}{0pt}{0}{49}{57}\ \genfrac{}{}{0pt}{0}{65}{152}\ \genfrac{}{}{0pt}{0}{74}{95}\ \genfrac{}{}{0pt}{0}{84}{209}\ \genfrac{}{}{0pt}{0}{96}{38}\ \genfrac{}{}{0pt}{0}{98}{19}\ \genfrac{}{}{0pt}{0}{112}{19}\ \genfrac{}{}{0pt}{0}{114}{95}\ \genfrac{}{}{0pt}{0}{130}{38}\ \genfrac{}{}{0pt}{0}{144}{19}\ \genfrac{}{}{0pt}{0}{147}{38}\ \genfrac{}{}{0pt}{0}{148}{38}\ \genfrac{}{}{0pt}{0}{200}{76}$   
\\[9pt]   
&  
$\genfrac{}{}{0pt}{0}{228}{38}\ \genfrac{}{}{0pt}{0}{252}{38}\ \genfrac{}{}{0pt}{0}{259}{38}\ \genfrac{}{}{0pt}{0}{260}{38}\ \genfrac{}{}{0pt}{0}{351}{38}\ \genfrac{}{}{0pt}{0}{616}{38}\ \genfrac{}{}{0pt}{0}{702}{19}\ \genfrac{}{}{0pt}{0}{777}{19}\ \genfrac{}{}{0pt}{0}{780}{19}\ \genfrac{}{}{0pt}{0}{798}{19}\ \genfrac{}{}{0pt}{0}{800}{19}\ \genfrac{}{}{0pt}{0}{1081}{19}\ \genfrac{}{}{0pt}{0}{1897}{19}\ \genfrac{}{}{0pt}{0}{5842}{19}\ \genfrac{}{}{0pt}{0}{43721}{1}$    
\\[5pt] 
\hline 
&   
\\[-6pt]
20
&  
$\genfrac{}{}{0pt}{0}{1}{2091}\ \genfrac{}{}{0pt}{0}{2}{2240}\ \genfrac{}{}{0pt}{0}{4}{2340}\ \genfrac{}{}{0pt}{0}{7}{1040}\ \genfrac{}{}{0pt}{0}{8}{1040}\ \genfrac{}{}{0pt}{0}{12}{720}\ \genfrac{}{}{0pt}{0}{14}{640}\ \genfrac{}{}{0pt}{0}{16}{505}\ \genfrac{}{}{0pt}{0}{21}{480}\ \genfrac{}{}{0pt}{0}{24}{360}\ \genfrac{}{}{0pt}{0}{28}{420}\ \genfrac{}{}{0pt}{0}{32}{60}\ \genfrac{}{}{0pt}{0}{37}{340}\ \genfrac{}{}{0pt}{0}{42}{260}\ \genfrac{}{}{0pt}{0}{48}{320}\ \genfrac{}{}{0pt}{0}{49}{130}\ \genfrac{}{}{0pt}{0}{56}{120}\ \genfrac{}{}{0pt}{0}{64}{20}\ \genfrac{}{}{0pt}{0}{65}{220}\ \genfrac{}{}{0pt}{0}{74}{120}\ \genfrac{}{}{0pt}{0}{84}{240}\ \genfrac{}{}{0pt}{0}{114}{160}\ \genfrac{}{}{0pt}{0}{130}{100}\ \genfrac{}{}{0pt}{0}{144}{50}\ \genfrac{}{}{0pt}{0}{147}{100}\ \genfrac{}{}{0pt}{0}{148}{120}\ \genfrac{}{}{0pt}{0}{168}{80}$   
\\[9pt]   
&  
$\genfrac{}{}{0pt}{0}{192}{20}\ \genfrac{}{}{0pt}{0}{196}{20}\ \genfrac{}{}{0pt}{0}{200}{100}\ \genfrac{}{}{0pt}{0}{228}{40}\ \genfrac{}{}{0pt}{0}{252}{40}\ \genfrac{}{}{0pt}{0}{259}{40}\ \genfrac{}{}{0pt}{0}{260}{40}\ \genfrac{}{}{0pt}{0}{351}{80}\ \genfrac{}{}{0pt}{0}{400}{40}\ \genfrac{}{}{0pt}{0}{441}{20}\ \genfrac{}{}{0pt}{0}{444}{40}\ \genfrac{}{}{0pt}{0}{455}{40}\ \genfrac{}{}{0pt}{0}{456}{40}\ \genfrac{}{}{0pt}{0}{616}{40}\ \genfrac{}{}{0pt}{0}{1081}{40}\ \genfrac{}{}{0pt}{0}{1232}{20}\ \genfrac{}{}{0pt}{0}{1365}{20}\ \genfrac{}{}{0pt}{0}{1368}{20}\ \genfrac{}{}{0pt}{0}{1369}{10}\ \genfrac{}{}{0pt}{0}{1400}{20}\ \genfrac{}{}{0pt}{0}{1404}{20}\ \genfrac{}{}{0pt}{0}{1897}{20}\ \genfrac{}{}{0pt}{0}{3329}{20}\ \genfrac{}{}{0pt}{0}{10252}{20}\ \genfrac{}{}{0pt}{0}{76725}{1}$       
\\[5pt] 
\hline
&   
\\[-6pt]
21 
& 
$\genfrac{}{}{0pt}{0}{1}{3062}\ \genfrac{}{}{0pt}{0}{2}{3465}\ \genfrac{}{}{0pt}{0}{4}{3633}\ \genfrac{}{}{0pt}{0}{7}{1617}\ \genfrac{}{}{0pt}{0}{8}{1876}\ \genfrac{}{}{0pt}{0}{12}{1092}\ \genfrac{}{}{0pt}{0}{14}{1008}\ \genfrac{}{}{0pt}{0}{16}{756}\ \genfrac{}{}{0pt}{0}{21}{756}\ \genfrac{}{}{0pt}{0}{24}{672}\ \genfrac{}{}{0pt}{0}{28}{861}\ \genfrac{}{}{0pt}{0}{32}{210}\ \genfrac{}{}{0pt}{0}{37}{504}\ \genfrac{}{}{0pt}{0}{42}{378}\ \genfrac{}{}{0pt}{0}{48}{441}\ \genfrac{}{}{0pt}{0}{49}{189}\ \genfrac{}{}{0pt}{0}{56}{126}\ \genfrac{}{}{0pt}{0}{64}{21}\ \genfrac{}{}{0pt}{0}{65}{357}\ \genfrac{}{}{0pt}{0}{74}{273}\ \genfrac{}{}{0pt}{0}{84}{609}\ \genfrac{}{}{0pt}{0}{96}{126}\ \genfrac{}{}{0pt}{0}{98}{63}\ \genfrac{}{}{0pt}{0}{112}{63}\ \genfrac{}{}{0pt}{0}{114}{231}\ \genfrac{}{}{0pt}{0}{130}{126}$   
\\[9pt]   
&  
$\genfrac{}{}{0pt}{0}{144}{63}\ \genfrac{}{}{0pt}{0}{147}{126}\ \genfrac{}{}{0pt}{0}{148}{126}\ \genfrac{}{}{0pt}{0}{200}{168}\ \genfrac{}{}{0pt}{0}{228}{105}\ \genfrac{}{}{0pt}{0}{252}{105}\ \genfrac{}{}{0pt}{0}{259}{105}\ \genfrac{}{}{0pt}{0}{260}{126}\ \genfrac{}{}{0pt}{0}{288}{21}\ \genfrac{}{}{0pt}{0}{294}{42}\ \genfrac{}{}{0pt}{0}{296}{42}\ \genfrac{}{}{0pt}{0}{336}{63}\ \genfrac{}{}{0pt}{0}{343}{7}\ \genfrac{}{}{0pt}{0}{351}{105}\ \genfrac{}{}{0pt}{0}{400}{42}\ \genfrac{}{}{0pt}{0}{441}{21}\ \genfrac{}{}{0pt}{0}{444}{42}\ \genfrac{}{}{0pt}{0}{455}{42}\ \genfrac{}{}{0pt}{0}{456}{42}\ \genfrac{}{}{0pt}{0}{616}{84}\ \genfrac{}{}{0pt}{0}{702}{42}\ \genfrac{}{}{0pt}{0}{777}{42}\ \genfrac{}{}{0pt}{0}{780}{42}\ \genfrac{}{}{0pt}{0}{798}{42}\ \genfrac{}{}{0pt}{0}{800}{42}\ \genfrac{}{}{0pt}{0}{1081}{42}\ \genfrac{}{}{0pt}{0}{1897}{42}$   
\\[9pt]   
&  
$\genfrac{}{}{0pt}{0}{2162}{21}\ \genfrac{}{}{0pt}{0}{2394}{21}\ \genfrac{}{}{0pt}{0}{2400}{21}\ \genfrac{}{}{0pt}{0}{2405}{21}\ \genfrac{}{}{0pt}{0}{2457}{21}\ \genfrac{}{}{0pt}{0}{2464}{21}\ \genfrac{}{}{0pt}{0}{3329}{21}\ \genfrac{}{}{0pt}{0}{5842}{21}\ \genfrac{}{}{0pt}{0}{17991}{21}\ \genfrac{}{}{0pt}{0}{134643}{1}$    
\\[5pt] 
\hline 
\end{tabular} 
}   
\label{t1} 
\end{table*}



\begin{table*}
\caption{The total number of different values of degeneracy $D$ for each input length $n$.} 
{\scriptsize  
\begin{tabular} 
{ccccccccccccccc} 
\hline 
&
&
&   
&  
&
&
&
&   
&  
&
&
&
&   
&  
\\[-6pt] 
$n\ $  
& 
3
& 
4
&
5
&
6
&
7
& 
8  
&  
9
&
10
&
11
&
12
& 
13  
&  
14
&
15
&
16
\\[0pt]  
$D\ $    
&
2
&
2
&
3
&
3
&
4
&   
5
& 
6 
&
7
&
9
&
11
&  
13 
&  
17
&
19
&
25
\\[1pt] 
\hline  
&   
&  
&
&
&
&   
&  
&
&
&
&   
&  
&
&
\\[-6pt] 
$n\,$ 
&
17
&
18
&
19
&
20
&
21
&
22
&
23
&
24
&
25
&
26
&
27
&   
28
&  
29
&
30     
\\[0pt]  
$D\,$ 
&
29
&
36
&
43
&
52
&
63
&
75
&
90
&
108
&
128
&
153
&
181
&   
215
&  
253
&
300 
\\[0pt] 
\hline  
&   
&  
&
&
&
&   
&  
&
&
&
&   
&  
&
&
\\[-6pt] 
$n\,$ 
&
31 
&
32
&
33
&
34
&
35
&
36
&
37
&
38
&
39
& 
40
& 
41
&
42
& 
43
&
44
\\[0pt]  
$D\,$ 
&
351
&
415
& 
485  
&  
569
&
665
&
777
&
904
&
1054
&
1223   
&
1421
&
1645
&
1905
&
2200
&
2543
\\[0pt] 
\hline  
&   
&  
&
&
&
&   
&  
&
&
&
&   
&  
&
&
\\[-6pt] 
$n\,$ 
&
45
&
46
&
47
&
48
&
49
&
50
&
51
&   
52
&  
53
&
54
&
55 
&
56
&
57
&
58
\\[0pt]  
$D\,$ 
&
2929
&
3375
&
3879
&
4461
&
5114
&
5868
&
6716
&   
7686
&  
8782
&
10030
&
11437
&
13040
& 
14841 
&  
16888
\\[0pt] 
\hline  
&   
&  
&
&
&
&   
&  
&
&
&
&   
&  
&
&
\\[-6pt] 
$n\,$ 
&
59
&
60
&   
61
&  
62
&
63
&
64 
&
65
&
66
&
67
&
68
&
69
&
70
&
71
&
72
\\[0pt]  
$D\,$ 
&
19190
&
21799
&   
24727
&  
28043
&
31761
&
35960
&
40667
& 
45973  
&  
51913
&
58600
&
66080
&
74482
&
83876
&
94416  
\\[0pt] 
\hline  
&   
&  
&
&
&
&   
&  
&
&
&
&   
&  
&
&
\\[-6pt] 
$n\,$ 
&
73
&
74
&   
75
&  
76
&
77
&
78 
&
79
&
80
&
81
&
82
&
83
&
84
&
85
&
86
\\[0pt]  
$D\,$ 
&
106179
&
119365
&   
134072
&  
150524
&
168868
&
189358
&
212176
& 
237646     
&  
265977
&
297558
&
332666
&
371756
&
415165
&
463454
\\[0pt] 
\hline  
&   
&  
&
&
&
&   
&  
&
&
&
&   
&  
&
&
\\[-6pt] 
$n\,$ 
&
87
&
88
&   
89
&  
90
&
91
&
92 
&
93
&
94
&
95
&
96
&
97
&
98
&
99
&
100
\\[0pt]  
$D\,$ 
&
517032
&
576564
&   
642571
&  
715835  
&
796997
&
887002
&
986631
& 
1096998
&  
1219086
&
1354211
&
1503550
&
1668712 
&
1851106 
&
2052643   
\\[0pt] 
\hline  
&   
&  
&
&
&
&   
&  
&
&
&
&   
&  
&
&
\\[-6pt] 
$n\,$ 
&
101
&
102
&   
103
&  
104
&
105
&
106 
&
107
&
108
&
109
&
110
&
111
&
112
&
113
&
114
\\[0pt]  
$D\,$ 
&
2275056  
&
2520605 
&   
2791384  
&  
3090106  
&
3419284 
&
3782133  
&
4181719   
& 
4621841  
&  
5106175
&
5639272 
&
6225525   
&
6870327 
&
7578971 
&
8357846   
\\[0pt] 
\hline  
&   
&  
&
&
&
&   
&  
&
&
&
&   
&  
&
&
\\[-6pt] 
$n\,$
&
115
&
116
&
117
&
118
&
119
& 
120
&
&
&
&
&
&
&
&
\\[0pt]  
$D\,$ 
&
9213269
&
10152854
&
11184127
&
12316088
&
13557775   
&
14919808
&
&
&
&
&
&
&
&
\\[0pt] 
\hline  
\end{tabular} 
}   
\label{t2} 
\end{table*}

 
\end{document}